\documentclass[reprint,superscriptaddress,amsmath,amssymb,aps,prl]{revtex4-2}

\usepackage{graphicx}% Include figure files
\usepackage{amsmath,txfonts}
\usepackage{bm}
\usepackage[colorlinks=true,citecolor=blue,linkcolor=blue,urlcolor=blue,bookmarks=false]{hyperref}
\usepackage{CJK}

\newcommand{\sgn}{\text{sgn}}
\def\unit#1{\mathord{\thinspace\rm #1}}

\newcommand\identity{1\kern-0.25em\text{l}}

\begin{document}
\begin{CJK}{UTF8}{bsmi}

\title{Four-band effective square lattice model for Bernal-stacked bilayer graphene} %Revisiting the Aharonov-Bohm experiment on gate-controlled bilayer graphene: A four-band effective model}% Force line breaks with 

\author{Szu-Chao Chen (陳思超)}
\affiliation{Department of Electro-Optical Engineering, National Formosa University, Yunlin, Taiwan}

% \author{Ching-Hung Chiu (邱靖鈜)}
% \affiliation{Department of Physics, National Cheng Kung University, Tainan 70101, Taiwan}

%\author{Chen-Chun Tai (戴辰駿)}
%\affiliation{Department of Physics, National Cheng Kung University, Tainan 70101, Taiwan}

\author{Alina Mre\'nca-Kolasi\'nska}
\email{alina.mrenca@fis.agh.edu.pl}
\affiliation{AGH University, Faculty of Physics and Applied Computer Science, al. Mickiewicza 30, 30-059 Krakow, Poland}

\author{\linebreak Ming-Hao Liu (劉明豪)}
\email{minghao.liu@phys.ncku.edu.tw}
\affiliation{Department of Physics and Center of Quantum Frontiers of Research and Technology (QFort), National Cheng Kung University, Tainan 70101, Taiwan}

\date{\today}

\begin{abstract}
Bernal-stacked bilayer graphene (BLG) provides an ideal basis for gate-controlled, and free of etching, electronic devices. Theoretical modeling of realistic devices is an essential part of research, %and can insure better understanding of the physical phenomena observed experimentally. 
however, simulations of large-scale BLG devices continue to be extremely challenging. Micrometer-sized systems are predominantly beyond the reach of the commonly used atomistic tight-binding method, %suffers from high memory demand/computational burden, 
while other numerical approaches based on the two dimensional %continuum model/
Dirac equation are not straightforward to conduct due to the fermion doubling problem. 
Here we present an approach based on the continuum model, %which takes advantage of the fermion doubling through the description of both $K$ and $K'$ valleys.
unharmed by the fermion doubling. 
 %Alhough the discretization of the $K$ valley component results in douled copied states, these copies are equivalent to the $K'$ valley and are thus useful/needed for inclusion of both valleys in the simulations. 
The discretization of the BLG continuum Hamiltonian %for the $K$ valley 
leads to an effective four-band model, with both valleys built-in.
We demonstrate its performance with realistic, large-scale systems, and obtain results consistent with experiments and with the tight-binding model, over a broad range of magnetic field. 
\end{abstract}

\maketitle
\end{CJK}

%\section{\label{sec intro}Introduction}

Bilayer graphene (BLG) is a versatile platform for a variety of electronic devices. 
Recently, superconductivity has been demonstrated in magic-angle BLG \cite{Cao2018} as well as 
Bernal-stacked BLG \cite{Zhou2022}, while BLG-based moir\'e heterostructures have been evidenced to manifest unconventional ferroelectricity \cite{Zheng2020}. 
The natural, Bernal, or AB-stacked bilayer graphene, thanks to its band gap controllable by displacement field, can be used as a base for electronic %devices 
components like gate-defined quantum point contacts (QPCs) \cite{Overweg2018, Kraft2018, Overveg2018topo, Banszerus2020, Lee2020}, quantum dots \cite{Allen2012, Goossens2012, Eich2018, Kurzmann2019, Banszerus2020}, 
 cavities \cite{Seemann2023}, and topological channels \cite{Martin2008, San-Jose2009, Li2016}. 
Exploiting the gate-tunable gap also led to the demonstration of
transverse magnetic focusing of carriers between gate-defined QPCs \cite{Ingla-Aynes2023}, and fully-gate-controlled interferometers \cite{Iwakiri2022, Fu2023}.

\begin{figure}[b]
\includegraphics[width=0.9\columnwidth]{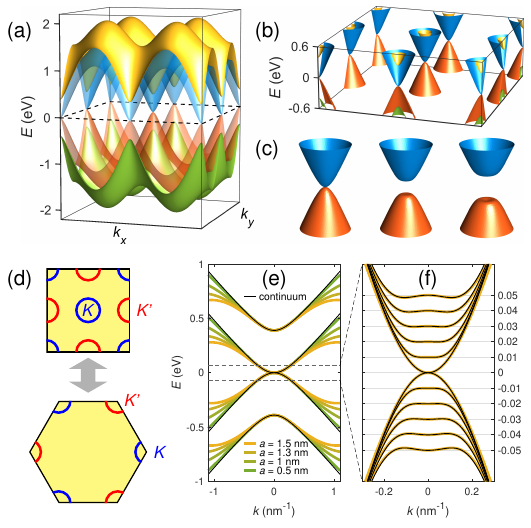} 
\caption{(a) Band structure obtained from the effective model. (b) Close-ups of the bands in (a) at low energy, showing exactly the same energy at the valleys at points $k_xa=0, \pm \pi,\ k_ya = 0, \pm \pi$. (c) Band structure at one of the valleys with $U=0,0.1,0.2\unit{eV}$. (d) Schematic diagram showing the mapping between the hexagonal and square lattice BZ with the $K$ ($K'$) valley marked by blue (red).
(e) Line cuts of the bands at $U=0$, with increasing square lattice spacing $a$, compared to the analytical energies (black curves). (f) The close-up marked in (e) with $a=1.5$ nm and $U$ varying from 0 to $0.1\unit{eV}$. Black curves are the analytical continuum model energies.} 
\label{fig:fig1}
\end{figure}

%%%%%%%% moved to the section about the ring %%%%%%%%
%Quantum rings represent the most basic realization of the Aharonov-Bohm two-slit experiment \cite{Aharonov1959} in solids. The magnetic flux piercing the area encircled by the two arms of the ring traversed by electrons introduces phase difference which leads to conductance oscillation with the period of magnetic quantum $\phi_0=h/e$. A number of works addressed this effect in quantum rings etched in graphene both theoretically \cite{Recher2007, Wurm2010, Mrenca2016} and experimentally \cite{Russo2008, Huefner2010, Smirnov2012, Smirnov2014, Dauber2017}. Inevitable consequences of their fabrication are edge roughness, disorder, and impurities which reduce the mean free path and coherence length. A better quality can be reached in interferometers relying on electrostatic confinement \cite{Ronen2021, Deprez2021, Iwakiri2023}, which naturally leads to the proposal of defining Aharonov-Bohm rings by inducing an energy gap in gated BLG \cite{Iwakiri2022}. 
%%A natural benefit of gate-controlled ring is a possibility to turn on and off the oscillation \cite{Iwakiri2022}. 
 %%%%%%%% %%%%%%%% %%%%%%%%

The theoretical modeling of quantum transport is commonly based on the atomistic tight-binding model. Modeling realistic Bernal-stacked BLG devices, of size of the order of hundreds of nanometers, is highly demanding, and faithful reproduction of experimental results was limited to rather simple geometries with lateral translational
invariance \cite{Varlet2014, Du2018}.
The scalable model for graphene \cite{Liu2015scalable} can be used to reduce the computational burden, however, in AB-stacked BLG the scaling factor required for reliable results is limited, making simulations of more complex devices challenging. Another approach is solving the Dirac equation within the continuum model, however, discretizing it may produce spurious solutions due to the fermion doubling. Circumventing this problem %is possible 
has been carried out by adding en extra dimension to the computational lattice \cite{Kaplan1992} or by various discretization schemes 
 \cite{Drell1976, Wilson1977, Susskind1977, Stacey1982, Kogut1983, Beenakker2023}. 
Many adaptations for condensed matter physics, and in particular graphene, have been developed \cite{Tworzydlo2008, Hernandez2012, Habib2016, Szafran2019, Ziesen2023} which in certain cases can also be extended to Bernal-stacked BLG \cite{Zebrowski2017}.

In this work, we show a discretization scheme of the continuum model which can capture transport properties of Bernal-stacked BLG. 
%present an effective model for Bernal-stacked BLG based on the continuum model, which  can tackle large, realistic devices and greatly reduce the computational burden. 
The resulting effective, four-band model can tackle large, realistic devices and greatly reduce the computational burden, as compared to the atomistic approaches.
In the discretization scheme derived for the $K$ valley, the fermion-doubled states show the properties of the $K$ or $K'$ valley, and thus we do not remove them from the spectrum with the methods developed before \cite{Zebrowski2017} since they are beneficial for simulating phenomena involving both valleys. 
%help model desired phenomena not included in a single-valley model. 
We demonstrate the advantage of this approach by recovering recent experiments in BLG: integer quantum Hall effect \cite{Novoselov2005, Zhang2005}, transverse magnetic focusing \cite{Taychatanapat2013}, Aharonov-Bohm effect in a gate-defined quantum ring \cite{Iwakiri2022}, and Klein tunneling in bipolar junctions \cite{Du2018}. 
%a recent experiment on a gate defined quantum ring in BLG \cite{Iwakiri2022}. 
%The Aharonov-Bohm oscillation obtained in BLG ring is confronted with a system of a similar geometry defined in single-layer graphene (SLG).

%\section{\label{sec methods}Methods}

%Methods go here
\textit{The four-band effective model.}
%\subsection{\label{sec model}The four-band effective model}
We begin our discussion with the continuum model Hamiltonian and  %its basic properties, and then proceed to 
the derivation of the effective model.
%The effective model is constructed by discretizing the four-orbital continuum model for bilayer graphene. The Hamiltonian is written in the basis of the orbitals on sites $A1,\ B1,\ B2,\ A2$, with lower (upper) layer labeled by 1 (2), and reads 
In the four-orbital continuum model for bilayer graphene, the Hamiltonian written in the basis of the orbitals on sites $A1,\ B1,\ B2,\ A2$, with lower (upper) layer labeled by 1 (2), reads \cite{McCann2013}
\begin{equation}
\begin{pmatrix}
V + U/2 	&	v_F \pi^+	&	\gamma_1	&	0	\\
v_F \pi^-	&	V + U/2		&	0			&	0	\\
\gamma_1	&	0			&	V - U/2		&	v_F \pi^- \\
	0		&	0			&	v_F \pi^+	&	V - U/2	\\
\end{pmatrix}\ ,
\label{eq:H_blg}
\end{equation}
where $v_F$ is the Fermi velocity of graphene, $U$ is the asymmetry parameter induced by the
electrical gating, $V$ is the band offset, $\gamma_1$ is the hopping integral between pairs of orbitals on the dimer sites $B1$ and $A2$, $\pi^\pm = \xi p_x \pm i p_y$, and the valley index $\xi=\pm$
for the $K$ $(K')$ valley. 
%The eigenstates are given by four-component wave function $\psi = \left(\psi_A^\mathrm{bot}, \psi_B^\mathrm{bot}, \psi_B^\mathrm{top}, \psi_B^\mathrm{top}\right)$. 
The discretization of the Hamiltonian on an artificial square lattice with four orbitals on each site $(A1,\ B1,\ B2,\ A2)$ is done using the central finite difference quotients for the spatial derivative. %the Kwant tool \texttt{discretize}.
Upon this procedure, we obtain the onsite energy term 
and hopping integrals, represented by $4\times 4$ matrices. The effective Hamiltonian is given by
\begin{equation}
\begin{aligned}
\sum\limits_{i,j} c^\dagger_{i,j} (V\identity + %\eta_{c_{i,j}} U/2)\identity
\frac{U}{2} \left[\tau_z\otimes \identity\right]) c_{i,j} 
+ \sum\limits_{i,j} \frac{1}{2} c^\dagger_{i,j} \left[
\tau_x \otimes 
% \left(
% \begin{smallmatrix} 
% \gamma_1&	0	\\
% 0		&	0	\\
% \end{smallmatrix} 
% \right)
\gamma_1(\sigma_z+\identity)
\right] c_{i,j} \\
+ \sum\limits_{i,j} \left( c^\dagger_{i+1,j} \frac{-it}{2} \left[\identity \otimes \sigma_x \right] c_{i,j} + H.c. \right) \\
+ \sum\limits_{i,j} \left( c^\dagger_{i,j+1} \frac{it}{2} \left[\tau_z\otimes \sigma_y\right] c_{i,j} + H.c. \right) 
\end{aligned}
\label{eq:H_eff}
\end{equation}
where $c^\dagger_{i,j} = (a^\dagger_{1,i,j}, b^\dagger_{1,i,j}, b^\dagger_{2,i,j}, a^\dagger_{2,i,j})$ and $c_{i,j} = (a_{1,i,j}, b_{1,i,j}, b_{2,i,j}, a_{2,i,j})$ are the creation and annihilation operators on a lattice site at $\mathbf{r}_{i,j}=(x_i, y_j)$ for an orbital on a sublattice $A$ or $B$ on the lower or upper layer, %$\eta_{c_i}=1$ for $c_{i,j}=a_{1,i,j}, b_{1,i,j}$, and $\eta_{c_i}=-1$ for $c_{i,j}=a_{2,i,j}, b_{2,i,j}$, 
$t=\hbar v_F/a$, $a$ is the square lattice spacing, $\sigma$ are the Pauli matrices operating on sublattice $A$ and $B$, and $\tau$ are the Pauli matrices operating on the two layers.

From the Hamiltonian \eqref{eq:H_eff} we derive the bulk BLG band structure
\begin{multline}
E = \pm \left( 
t^2 (\sin^2 k_xa + \sin^2 k_ya) + \frac{U^2}{4} + \frac{\gamma_1^2}{2} \right. \\
\left. \pm \frac{1}{2}
\left[ 
\gamma_1^4 + 4 t^2 (\sin^2 k_xa + \sin^2 k_ya) (U^2 + \gamma_1^2)
\right]^{1/2}
\right)^{1/2};
\label{eq:E_eff_2d}
\end{multline}
see Supplemental Material (SM).
The four energy bands obtained from this model for $U=0$ are presented in \autoref{fig:fig1}(a).
 One can see that in the first Brillouin zone (BZ), the Dirac cone is copied to the points $k_xa=0, \pm \pi,\ k_ya = 0, \pm \pi$ [\autoref{fig:fig1}(b)]. %The additional Dirac cones emerge due to the fermion doubling. , which generates $2^d$-fold degeneracy in the system of dimension $d$. 
 The low energy band structure considering a single cone and $U=0,\ 0.1,\ 0.2$ eV is shown in \autoref{fig:fig1}(c).

To determine which valley a cone corresponds to, we perform a low-$k$ expansion around a given $k$-point (see SM). %\textcolor{red}{(see SM)}. 
We find that in the effective model, there are in total two $K$ and two $K'$ valleys in the first BZ [\autoref{fig:fig1}(d), upper panel], while in the tight binding model, the first BZ contains two nonequivalent valleys $K$ and $K'$ [\autoref{fig:fig1}(d), lower panel]. Thus the effective model, although obtained by discretizing the continuum model for the $K$ valley, can capture the properties of both $K$ and $K'$ valleys. Bearing in mind that in the transport calculations the resulting conductance will also be doubled, division of the total conductance by 2 is sufficient to correct this doubling. 
%In the tight binding model, the first BZ contains two nonequivalent valleys $K$ and $K'$ [\autoref{fig:fig1}(d), lower panel]. 
%On the other hand, in the effective model, there are in total two $K$ and two $K'$ valleys in the first BZ [\autoref{fig:fig1}(d), upper panel], thus the  in the transport calculations the total conductance is also doubled, and division by 2 is used to correct for this doubling.

\autoref{fig:fig1}(e) shows the line cuts of the band structure given by the analytical formula from the continuum model and by (\autoref{eq:E_eff_2d}) around the cone in the center of the BZ, with a set of $a$ values and $U=0$. At low energy the two formulas are consistent. Upon increasing $a$, (\autoref{eq:E_eff_2d}) starts to deviate from the exact band structure at lower $k$. The low-energy discretized model (\autoref{eq:E_eff_2d}) also agrees with the continuum model with $U\ne 0$ [\autoref{fig:fig1}(f), shown with $a=1.5$ nm, the largest of the spacings in \autoref{fig:fig1}(e)]. The model catches the characteristic 'Mexican hat' structure and is accurate close to the band edge even at the asymmetry parameter reaching $U=0.1$ eV.
Moreover, (\autoref{eq:E_eff_2d}) is to a good approximation isotropic at low energy. This property is used for the calculation of the band offset for the transport calculations (see SM). %\textcolor{red}{(see SM)}.

\begin{figure}[t]
    \centering
    \includegraphics{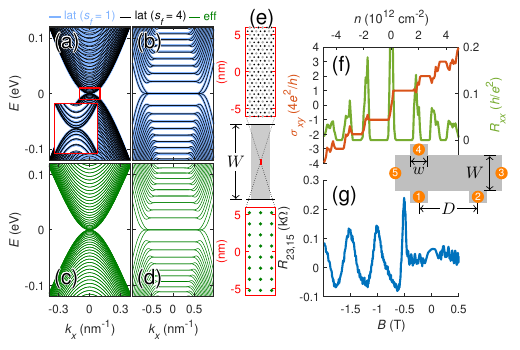}
    \caption{Band structures of a nanoribbon (a) -- (b)  of width $W=200$ nm, calculated using the tight binding model, and (c)-- (d) of width $W=201$ nm, calculated using the effective model. In (a) and (c) $B=0$, and in (b) and (d) $B=5$ T. (e) middle panel: sketch of the ribbon, upper (lower) panel: zoom-in showing the structure for tight binding (effective model) calculation. (f) Longitudinal resistance $R_{xx}$ and Hall conductivity $\sigma_{xy}$ as a function of density, at $B=14$ T, calculated with grid spacing $a=1$ nm. Inset: sketch of the system for the transport calculation. (g) Nonlocal resistance $R_{23,15}$ as a function of $B$, at $n=-2.8\times 10^{12}$ cm$^{-2}$.}
    \label{fig:fig2new}
\end{figure}

\textit{Transport calculation.}
Before we proceed to the applications% and comparison of the tight binding and the effective model results
, let us first mention the transport calculation methods. 
%\subsection{\label{sec transport}Transport calculation}
The calculations for BLG are based on the effective model, as well as the tight-binding model. %, whereas for SLG we use only the latter. 
The tight-binding Hamiltonian can be summarized as
\begin{equation}
\label{eq:Htb}
H = -\sum\limits_{\left\langle {i,j} \right\rangle } t_{ij} c_i^\dagger  c_j +\sum\limits_j {\varepsilon({\mathbf{r}}_j)} c_j^\dagger c_j,
\end{equation}
where the first sum runs over nearest neighbors, and the second sum contains the onsite energies, $c_i$ ($c_i^\dag$) is annihilation (creation) operator of an electron on site $i$ with the coordinates $\mathbf{r}_i=(x_i,y_i)$, 
%In SLG, 
$t_{ij}$ is the nearest-neighbor hopping parameter with $t_{ij}=t_0=3$ eV, and $\varepsilon$ is the on-site energy. 
In addition, we adopt the scalable tight-binding model \cite{Liu2015scalable}, where the hopping parameter is scaled as $t' = t_0/s_\mathrm{F}$ and the lattice spacing $a' = a_0 s_\mathrm{F}$ , $s_\mathrm{F}$ is the scaling factor, and we use 
$a=0.142$ nm. %$a_0 =0.246/\sqrt{3}$ nm. 
In BLG, in addition to the intralayer $t_0$, the interlayer hoppings between the dimer sites are included, with $t_{ij}=\gamma_1=0.39$ eV.
The external magnetic field $\mathbf{B}=(0,0,B)$ is introduced by including the Peierls phase in the hopping integrals, $t_{ij} \rightarrow t_{ij} \mathrm{e}^{i\phi}$, with $\phi = -\tfrac{e}{\hbar} \int_{\textbf{r}_i}^{\textbf{r}_j} \textbf{A}\cdot d\textbf{r}$, where the integration runs from the site at $\textbf{r}_i$ to the site at $\textbf{r}_j$, and the vector potential $\mathbf{A}$ satisfies $\nabla\times \mathbf{A} =\mathbf{B}$. %in the Landau gauge $\textbf{A}=(-yB, 0, 0)$. %\textcolor{red}{depends if the systems are perpendicular or vertical...}. 
%For this choice of the gauge, 
In the effective model, the magnetic field is also introduced via the Peierls substitution $t \rightarrow t \mathrm{e}^{i\phi}$. %which satisfies $\nabla\times \mathbf{A} = \mathbf{B}$.
The transport calculation is performed using the Kwant package \cite{Groth2014} for the nanoring, and the wave function matching method \cite{Kol2016transport} for the other systems.
%At zero temperature the conductance from lead $i$ to lead $j$ is calculated using the Landauer formula $G_{ji}=2e^2T_{ji}/h $, where $T=\sum_m T_{ji}^{(m)}$ is summed over the propagating modes.

\textit{Nanoribbons and multiterminal transport.}
Having discussed the properties of the effective model in bulk BLG and the transport calculation methodology, we demonstrate that the model applies well to finite width graphene ribbons, which are typically used as leads in two- and multi-terminal transport geometries. In \autoref{fig:fig2new}(a) -- \autoref{fig:fig2new}(e) we consider a nanoribbon $W\approx 200$ nm wide [\autoref{fig:fig2new}(e), middle panel]. The zoom-in of the structure used in the tight-binding (effective) model calculation is shown in \autoref{fig:fig2new}(e) upper (lower) panel. 
Band structures of a 200 nm ribbon calculated within the tight binding model with scaling factor $s_F=1$ and its scaled variant with $s_F=4$ are presented in \autoref{fig:fig2new}(a), and the one for a 201 nm wide ribbon obtained with the effective model ($a=1.5$ nm) -- in \autoref{fig:fig2new}(c). The low-energy band structure remains in a good agreement within both approaches [see low-energy zoom-in in the inset of \autoref{fig:fig2new}(a)]. The agreement remains equally good at finite magnetic field, as shown in \autoref{fig:fig2new}(b) and \autoref{fig:fig2new}(d), at $B=5$ T. 

Next, to demonstrate the model performance at moderate to high magnetic field, we revisit the experiments on transverse magnetic focusing (TMF) \cite{Taychatanapat2013} %at moderate $B$ field, 
and integer quantum Hall effect \cite{Novoselov2005}. %at high magnetic field. 
The multiterminal transport calculation was done within the Landauer-B\"uttiker approach, as described in 
Ref.\ \cite{Mrenca2023}. Generally, we consider the four-probe resistance $R_{ij,kl}$ with current flowing between $i$ and $j$ probes, and voltage between $k$ and $l$ probes. %The results obtained within the effective model with grid spacing $a=1.5$ nm are presented in \autoref{fig:fig2new}(f) and \autoref{fig:fig2new}(g). 
For both simulations, we used a 5-terminal geometry shown in the inset between \autoref{fig:fig2new}(f) and \autoref{fig:fig2new}(g), with the dimensions varying for both cases. 

\autoref{fig:fig2new}(f) shows the Hall conductivity $\sigma_{xy}$ and the longitudinal resistance $R_{xx}=R_{53,12}$ [see the probe labels in \autoref{fig:fig2new}(f)] at $B=14$ T, considering the probe spacing $D=401$ nm, width $w=199$ nm, and scattering region width $W=268$ nm, with $a=1$ nm. To induce the Landau level broadening, in the scattering region (but not in leads) we add a random onsite potential with a uniform distribution within the range $\pm 0.1$ eV. The calculated $\sigma_{xy}$ shows plateaus quantized at $\pm (4 e^2/h) N,\ N=1,2,...$, with $R_{xx}$ assuming nonzero value in between the plateaus and zero otherwise. Within the presented density range, up to the 4th plateau develops, in agreement with the experiment \cite{Novoselov2005}. At high density, the model starts to show small discrepancy, namely doubling of the $R_{xx}$ peaks as a result of the subbands 
 protruding below the flat subbands in the band structure [\autoref{fig:fig2new}(d)]. %the differences between both models in the nanoribbon band structure at energies further from the charge neutrality point.

For the TMF calculation, we consider a clean system with $D=774$ nm, $w=100.5$ nm, and $W=567$ nm, with $a=1.5$ nm. \autoref{fig:fig2new}(g) shows the nonlocal resistance $R_{23,15}$ calculated at $n=-2.8\times 10^{12}$ cm$^{-2}$, as a function of magnetic field. We observe peaks of nonlocal resistance roughly at multiples of $B_1 \approx 0.5$ T, in a good agreement with Ref.\ \cite{Taychatanapat2013}, and corresponding to cyclotron radius $R_\mathrm{c} = \hbar\sqrt{\pi n}/eB_1 \approx 390$ nm. To sum up, we demonstrated that the model works well for realistic large-scale systems.

\textit{The Aharonov-Bohm effect.}
%\subsection{\label{sec onsite}The gated ring model}
Quantum rings represent the most basic realization of the Aharonov-Bohm two-slit experiment \cite{Aharonov1959} in solids. The magnetic flux piercing the area encircled by the two arms of the ring traversed by electrons introduces phase difference which leads to conductance oscillation with the period of magnetic flux quantum $\phi_0=h/e$. A number of works addressed this effect in quantum rings etched in graphene both theoretically \cite{Recher2007, Wurm2010, Mrenca2016} and experimentally \cite{Russo2008, Huefner2010, Smirnov2012, Smirnov2014, Dauber2017}. Inevitable consequences of their fabrication are edge roughness, disorder, and impurities which reduce the mean free path and coherence length. A better quality can be reached in interferometers relying on electrostatic confinement \cite{Ronen2021, Deprez2021, Iwakiri2023}, which naturally leads to the proposal of defining Aharonov-Bohm rings by inducing an energy gap in gated BLG \cite{Iwakiri2022} revisited below. 

\begin{figure}[t]
\includegraphics[width=\columnwidth]%, trim={1.5cm 0 0.6cm 0},clip]
{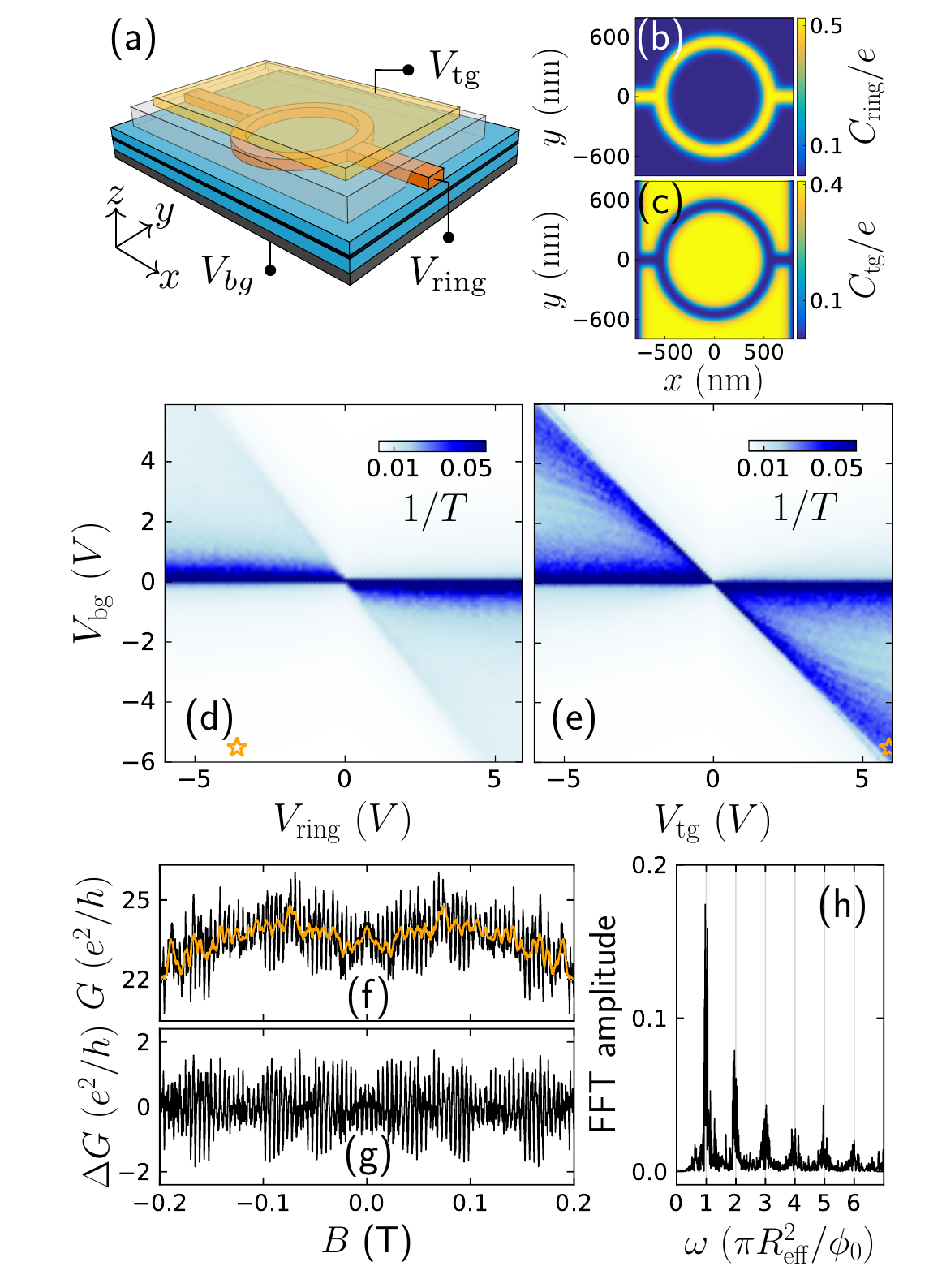} % l d r t
\caption{(a) Scheme of the considered system. (b)--(c) Capacitance profile used for modeling the ring gate (b) and top gate (c). (d) Inverse transmission $1/T$ as a function of $V_\mathrm{bg}$ and $V_\mathrm{ring}$ with $V_\mathrm{tg}=0$, and (e) as a function of $V_\mathrm{bg}$ and $V_\mathrm{tg}$ with $V_\mathrm{ring}=0$. 
(f) Conductance as a function of magnetic field at $V_\mathrm{bg} = -5.5152\ \mathrm{V},\ V_\mathrm{tg} = 5.8788\ \mathrm{V},\ V_\mathrm{ring}=-3.6\ \mathrm{V}$ (black curve) and smooth background (orange curve). (g) Oscillatory part $\Delta G$ and (h) its Fourier transform.
%\textcolor{red}{Mark the radii in (b), the system sizes in (c).}
} 
\label{fig:fig2}
\end{figure}

The system studied here is presented in \autoref{fig:fig2}(a). BLG (black layer in \autoref{fig:fig2}(a)) is encapsulated in hBN (blue) of thickness $d_\mathrm{b}=45.1$ nm for the lower hBN layer, and $d_\mathrm{t}=37.3$ nm for the upper hBN layer. The hBN/BLG/hBN sandwich is placed on a global back gate (dark gray), which tunes the band offset and asymmetry parameter together with the ring-shaped gate (orange), and the top gate (yellow semi-transparent). The top gate is separated from the ring gate by a layer of aluminum oxide (white semi-transparent). 
The back gate capacitance is obtained from the parallel-plate capacitor model as $C_\mathrm{bg}/e = \varepsilon_0\varepsilon_\mathrm{hBN}/e d_\mathrm{b} = 0.40437 \times 10^{12}$ cm$^{-2}$V$^{-1}$, where $\varepsilon_0$ is the vacuum permittivity, and $-e$ is the electron charge. 
For the modeling of the ring-shaped gate, we use a model function $C_\mathrm{ring}(x,y)$, and for the top gate, $C_\mathrm{tg}(x,y)$, shown in \autoref{fig:fig2}(b) and \autoref{fig:fig2}(c), respectively (see SM). 
The carrier density induced by the gates is given by 
\begin{equation}
\label{eq:dens}
n = (C_\mathrm{bg}V_\mathrm{bg} + C_\mathrm{ring}V_\mathrm{ring} + C_\mathrm{tg}V_\mathrm{tg}) /e
\end{equation} 
when there is no intrinsic doping. Here, %$C_\mathrm{ring}=C_\mathrm{ring}(x,y)$ and $C_\mathrm{tg}=C_\mathrm{tg}(x,y)$ thus the density is position-dependent. 
$n=n(x,y)$ is position-dependent. 
From the band structure of the effective model (\autoref{eq:E_eff_2d}), given the asymmetry parameter $U$ and the carrier density $n$, we get the band offset $V$ (as described in SM). 

% In the calculations for single-layer graphene, we consider the same device design as for BLG, the only difference being the onsite energies. For single layer graphene, it is given by
% \begin{equation}
% \label{eq:SLGE}
% \varepsilon=-\sgn(n) \hbar v_F \sqrt{\pi |n|}.
% \end{equation}

%\section{\label{sec results}Results}
%\subsection{\label{sec blg} Bilayer graphene ring}
%For the transport calculation w
We consider a system %of width $W$ and length $L$ \textcolor{red}{[see \autoref{fig:fig2}(c)]}, 
connected to two contacts simulated as semi-infinite leads. %[see \autoref{fig:fig2}(c)]. 
In the system described above it is possible to tune independently the densities and band gaps in the region under the ring-shaped gate and the region beyond it but covered by the top gate (referred to as bulk from now on). To form a ring, BLG in the ring region is tuned to a non-zero density while the bulk should be set within the bandgap.
To find the conditions in which these requirements are satisfied, we first calculate the inverse transmission $1/T$ at zero magnetic field with either $V_\mathrm{ring}=0$ or $V_\mathrm{tg}=0$, and sweeping the remaining voltages. For the calculation within the effective model we choose zero Fermi energy and lattice spacing $a=1.5$ nm. \autoref{fig:fig2}(d) shows $1/T$ as a function of $V_\mathrm{ring}$ and $V_\mathrm{bg}$, with $V_\mathrm{tg}=0$. The nearly horizontal line of high $1/T$ is the charge neutrality line of the region over the backgate and not covered by the top gate and ring gate. The diagonal line of slightly increased $1/T$ between $(V_\mathrm{ring}, V_\mathrm{bg}) = (-5\ \mathrm{V}, 6\ \mathrm{V})$ and $(5\ \mathrm{V}, -6\ \mathrm{V})$ %, marking the boundary between a finite $1/T$ and near zero one, 
is the charge neutrality line of the region controlled by the ring gate. %Between this line and the $V_\mathrm{bg}\approx 0$ one, the ring-shaped region is conducting and has a finite carrier density. 
Similarly, in \autoref{fig:fig2}(e), the scan of $1/T$ is plotted as a function of $V_\mathrm{tg}$ and $V_\mathrm{bg}$, with $V_\mathrm{ring}=0$. We observe a similar behavior but with a slightly lower slope of the bulk charge neutrality line from $(V_\mathrm{tg}, V_\mathrm{bg}) = (-6\ \mathrm{V}, 6\ \mathrm{V})$ and $(6\ \mathrm{V}, -6\ \mathrm{V})$. Choosing a combination of $(V_\mathrm{bg}, V_\mathrm{tg})$ along this charge neutrality line, one can gap out the bulk region, while the ring-shaped area is conducting. %In the following, we focus on the parameters $V_\mathrm{bg} = -5.5152\ \mathrm{V},\ V_\mathrm{tg} = 5.8788\ \mathrm{V},\ V_\mathrm{ring}=-3.6\ \mathrm{V}$ (\textcolor{red}{marked by dots in \autoref{fig:fig2}...}).

Comparison of  $1/T$ calculated here and the two-terminal resistance in Ref.\ \cite{Iwakiri2022} deserves a comment. The diagonal charge neutrality line in the scan of $1/T(V_\mathrm{ring}, V_\mathrm{bg})$  is less pronounced in the calculations than in the measurement \cite{Iwakiri2022} due to the presence of conducting region in the bulk area. %The transmission through the system is suppressed along the backgate charge neutrality line (when the region between the lead and the top gate is cut off(?)). 
Moreover, the experiment \cite{Iwakiri2022} showed two horizontal charge neutrality lines, likely caused by nonuniform intrinsic doping in certain regions of the device. The effects of intrinsic doping were neglected here since the main focus of this study is the quantum ring performance. Despite these qualitative differences, the agreement between the $1/T$ maps and the two-terminal resistance in Ref. \cite{Iwakiri2022} is good, particularly the positions of the diagonal charge neutrality lines.

Next, we can proceed to the magnetotransport calculation for the selected gate voltage values $V_\mathrm{bg} = -5.5152\ \mathrm{V},\ V_\mathrm{tg} = 5.8788\ \mathrm{V},\ V_\mathrm{ring}=-3.6\ \mathrm{V}$, [marked by stars in \autoref{fig:fig2}(d) and \autoref{fig:fig2}(e)], 
with the ring forms in the device. 
For the nominal dimensions of the ring, with $R_\mathrm{in}=500$ nm and $R_\mathrm{out}=580$ nm being the inner and outer radius of the ring, respectively, we expect the Aharonov-Bohm oscillation period to be within the range $2\hbar/eR_\mathrm{out}^2=3.9$ mT to $2\hbar/eR_\mathrm{in}^2=5.26$ mT. 
\autoref{fig:fig2}(f) shows the conductance calculated for magnetic field between $\pm 200$ mT (black line), while the fluctuating orange line is the smooth background. The background is obtained by smoothing the curve with Savitzky-Golay filter, using a window size 7.5 mT and polynomial order 2. It is then subtracted from the original curve to show the oscillatory part of the conductance $\Delta G$ in \autoref{fig:fig2}(g). By performing fast Fourier transform (FFT) of $\Delta G$, we obtain the spectrum shown in \autoref{fig:fig2}(h), where the frequency axis is presented in the units of $\pi R_\mathrm{eff}/\phi_0$, and $R_\mathrm{eff}$ is the effective radius of the ring. % estimated from FFT signal to be 550 nm. 
The FFT spectrum contains a set of peaks falling at $\upsilon\times \pi R_\mathrm{eff}/\phi_0$ which allows to estimate $R_\mathrm{eff}=550$ nm that gives $\upsilon$ close to integer values [\autoref{fig:fig2}(h)]. The first and most pronounced peak corresponds to a single magnetic flux piercing the area of the ring with an effective radius $R_\mathrm{eff}$, and the AB period $4.35$ mT. The further peaks represent the higher harmonics at multiples of magnetic flux through the ring area, as the carrier paths encircling the ring multiple times in the clockwise or anticlockwise direction accumulates a phase difference proportional to multiples of $\phi_0$. %interference of the wave function of a charge carrier encircling the ring multiple times. 
The probability for each subsequent $n$th round decreases which leads to the decrease of the overall amplitude of the $n$th peak. The damping of the peaks amplitude is usually considered as due to phase breaking \cite{Hansen2001}, however the present results are obtained in a fully coherent calculation. Thus we conclude that the decrease of amplitude of higher peaks is due to the carrier wave function leaving the ring in parts in each subsequent round. %, while the other part circulates in the ring. 

\textit{Fabry-P\'erot interference.}
To directly compare the performance of the effective model to the tight-binding model, we consider a Fabry-P\'erot interferometer in BLG. The calculation is based on the dual-gated device described in Ref. \cite{Du2018}, with the top gate width 157 nm. For the transport calculation, we assume the system is translationally invariant in the lateral direction, and we use the method of periodic hopping \cite{Wimmer2008, 2012periodicModel, Liu2012periodic}, so that we need to consider the $x$-dependence only of the carrier density. For the simplicity of comparison, we present the normalized conductance calculated as $g=(e^2/h)\int_{-k_{\mathrm{F}}}^{k_{\mathrm{F}}}T(k_y)d k_y$, with $k_{\mathrm{F}}$ being the Fermi momentum. 
To calculate the density profile along the system as $n(x) = (V_\mathrm{tg}C_\mathrm{tg}+V_\mathrm{bg}C_\mathrm{bg}+V_\mathrm{c}C_\mathrm{c})/e$, we use the capacitance for the bottom gate $C_\mathrm{bg}(x)$, top gate $C_\mathrm{tg}(x)$, and contacts $C_\mathrm{c}(x)$ obtained from the finite element electrostatic simulation (for details on the geometry and electrostatic modeling see Ref. \cite{Du2018}).

\begin{figure}[t]
\includegraphics[width=\columnwidth]{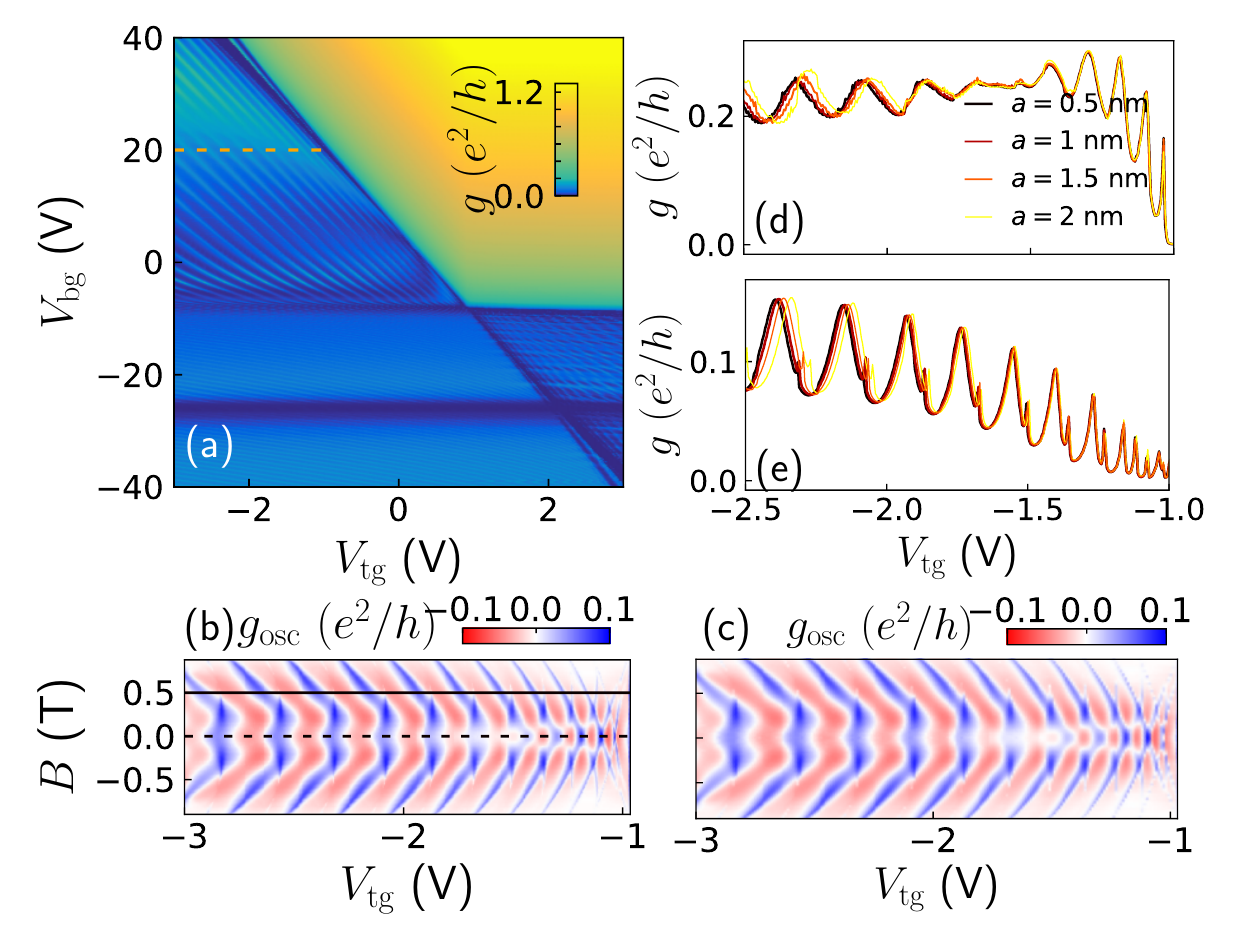} 
\caption{(a) Conductance in dual-gated device as a function of $V_\mathrm{tg}$ and $V_\mathrm{bg}$ at zero magnetic field obtained within the effective model with $a=1$ nm. (b) Effective model and (c) tight-binding model simulation of conductance with varying magnetic field and top gate voltage along the dashed line in (a). (d-e) Conductance line cuts at (d) $B=0$ and (e) $B=0.4$ T,  marked in (b) with dashed and solid line, respectively.}
\label{fig:fig4}
\end{figure}

\autoref{fig:fig4}(a) shows the conductance as a function of $V_\mathrm{tg}$ and $V_\mathrm{bg}$, with $V_\mathrm{c}=1.1$ V, calculated with the effective four-band model with the grid spacing $a=1$ nm. The map is consistent with the simulation result in Ref. \cite{Du2018}. For a demonstration of the performance of the effective model at finite magnetic field, we perform magnetotransport calculations. We choose a range of $V_\mathrm{tg}$ at $V_\mathrm{bg}=20$ V, marked with the orange dashed line in \autoref{fig:fig4}(a), and investigate the weak $B$ dependence of conductance. \autoref{fig:fig4}(b) shows the oscillating part of conductance $g_\mathrm{osc}$ as a function of $V_\mathrm{tg}$ and $B$ calculated using the effective model with $a=0.5$ nm, and \autoref{fig:fig4}(c) with the tight binding model. %with the scaling factor $s_F=1$. 
$g_\mathrm{osc}$ was obtained by subtracting the smooth background from the overall conductance $g$. %obtained by Savitzky-Golay filter with polynomial order 2 and window size 0.5 V.
The effective model reproduces faithfully the tight binding model results, including the oscillation period, the $\pi$-phase shift at finite magnetic field at $V_\mathrm{tg}>-1.5$ V, the transition to no phase shift below that voltage, and parabolic dispersion of the fringes with magnetic field. %With grid spacing $a=1$ nm the size of the numerical problem is already significantly reduced with respect to the tight binding model, but 
We further check the influence of the $a$ value. \autoref{fig:fig4}(d) and \autoref{fig:fig4}(e) show the line cuts of conductance at $B=0$ and $B=0.4$ T, respectively, %, as marked in  with dashed and solid line.
with grid spacing varying from 0.5 nm to 2 nm. For increasing $a$, the results are consistent at low $V_\mathrm{tg}$, but start to deviate slightly 
for higher voltage, far from the charge neutrality point, particularly at $a=2$ nm. The overall qualitative agreement is very good. We conclude that $a=1.5$ nm, also used in the gated ring simulation above, is satisfactorily precise, and the size of the numerical problem is already significantly reduced with respect to the tight binding model.

%\section{Concluding remarks}

\textit{Concluding remarks.}
We developed a method for simulation of realistic BLG devices of micrometer size, which can reduce the calculation time and memory requirements.
This method can be particularly useful for large-scale systems like BLG superlattices, which have recently been proposed as a medium for flat energy bands \cite{Ghorashi2023, Sun2023}. Our modeling can be used for description of other experimental devices which have been beyond the reach of tight-binding modeling, such as quantum point contacts with constricton defined by displacement field \cite{Kraft2018, Overveg2018topo} as well as electrostatic BLG quantum dots \cite{Eich2018, Kurzmann2019, Banszerus2018, Banszerus2020, Banszerus2020dbl}.
 Although the discretized continuum Hamiltonian is a simple one with only dimer interlayer hopping, that does not contain the trigonal warping  %obtained when including
requiring inclusion of the skew interlayer hopping, it is sufficient for basic applications presented in this paper, and generally for the description of phenomena which probe the energy range beyond the trigonally warped region. %charge neutrality point. 
The trigonal warping effects can manifest themselves in certain transport experiments \cite{Gold2021}. For such specific applications, the skew hopping could be included in the original continuum Hamiltonian (\autoref{eq:H_blg}), which would yield a modified effective model. 
As shown above, the effective four-band model accounts for both valleys $K$ and $K'$, similarly to the tight-binding model. Such correspondence can be extended to other hexagonal materials, providing a tool for efficient modeling of realistic devices.  

\begin{acknowledgments}
We thank Ching-Hung Chiu and Chen-Chun Tai for useful discussions.
Financial supports from National Science and Technology Council of Taiwan (grant numbers:  MOST 109-2112-M-006-020-MY3 and NSTC 112-2112-M-150-001-MY3) is gratefully acknowledged. This
research was supported in part by program ,,Excellence initiative -- research university'' for the AGH University of Krakow and by PL-Grid Infrastructure.
\end{acknowledgments}

\bibliographystyle{apsrev4-1}
\bibliography{AB}

\begin{thebibliography}{63}%
\makeatletter
\providecommand \@ifxundefined [1]{%
 \@ifx{#1\undefined}
}%
\providecommand \@ifnum [1]{%
 \ifnum #1\expandafter \@firstoftwo
 \else \expandafter \@secondoftwo
 \fi
}%
\providecommand \@ifx [1]{%
 \ifx #1\expandafter \@firstoftwo
 \else \expandafter \@secondoftwo
 \fi
}%
\providecommand \natexlab [1]{#1}%
\providecommand \enquote  [1]{``#1''}%
\providecommand \bibnamefont  [1]{#1}%
\providecommand \bibfnamefont [1]{#1}%
\providecommand \citenamefont [1]{#1}%
\providecommand \href@noop [0]{\@secondoftwo}%
\providecommand \href [0]{\begingroup \@sanitize@url \@href}%
\providecommand \@href[1]{\@@startlink{#1}\@@href}%
\providecommand \@@href[1]{\endgroup#1\@@endlink}%
\providecommand \@sanitize@url [0]{\catcode `\\12\catcode `\$12\catcode
  `\&12\catcode `\#12\catcode `\^12\catcode `\_12\catcode `\%12\relax}%
\providecommand \@@startlink[1]{}%
\providecommand \@@endlink[0]{}%
\providecommand \url  [0]{\begingroup\@sanitize@url \@url }%
\providecommand \@url [1]{\endgroup\@href {#1}{\urlprefix }}%
\providecommand \urlprefix  [0]{URL }%
\providecommand \Eprint [0]{\href }%
\providecommand \doibase [0]{http://dx.doi.org/}%
\providecommand \selectlanguage [0]{\@gobble}%
\providecommand \bibinfo  [0]{\@secondoftwo}%
\providecommand \bibfield  [0]{\@secondoftwo}%
\providecommand \translation [1]{[#1]}%
\providecommand \BibitemOpen [0]{}%
\providecommand \bibitemStop [0]{}%
\providecommand \bibitemNoStop [0]{.\EOS\space}%
\providecommand \EOS [0]{\spacefactor3000\relax}%
\providecommand \BibitemShut  [1]{\csname bibitem#1\endcsname}%
\let\auto@bib@innerbib\@empty
%</preamble>
\bibitem [{\citenamefont {Cao}\ \emph {et~al.}(2018)\citenamefont {Cao},
  \citenamefont {Fatemi}, \citenamefont {Fang}, \citenamefont {Watanabe},
  \citenamefont {Taniguchi}, \citenamefont {Kaxiras},\ and\ \citenamefont
  {Jarillo-Herrero}}]{Cao2018}%
  \BibitemOpen
  \bibfield  {author} {\bibinfo {author} {\bibfnamefont {Y.}~\bibnamefont
  {Cao}}, \bibinfo {author} {\bibfnamefont {V.}~\bibnamefont {Fatemi}},
  \bibinfo {author} {\bibfnamefont {S.}~\bibnamefont {Fang}}, \bibinfo {author}
  {\bibfnamefont {K.}~\bibnamefont {Watanabe}}, \bibinfo {author}
  {\bibfnamefont {T.}~\bibnamefont {Taniguchi}}, \bibinfo {author}
  {\bibfnamefont {E.}~\bibnamefont {Kaxiras}}, \ and\ \bibinfo {author}
  {\bibfnamefont {P.}~\bibnamefont {Jarillo-Herrero}},\ }\href {\doibase
  10.1038/nature26160} {\bibfield  {journal} {\bibinfo  {journal} {Nature}\
  }\textbf {\bibinfo {volume} {556}},\ \bibinfo {pages} {43} (\bibinfo {year}
  {2018})}\BibitemShut {NoStop}%
\bibitem [{\citenamefont {Zhou}\ \emph {et~al.}(2022)\citenamefont {Zhou},
  \citenamefont {Holleis}, \citenamefont {Saito}, \citenamefont {Cohen},
  \citenamefont {Huynh}, \citenamefont {Patterson}, \citenamefont {Yang},
  \citenamefont {Taniguchi}, \citenamefont {Watanabe},\ and\ \citenamefont
  {Young}}]{Zhou2022}%
  \BibitemOpen
  \bibfield  {author} {\bibinfo {author} {\bibfnamefont {H.}~\bibnamefont
  {Zhou}}, \bibinfo {author} {\bibfnamefont {L.}~\bibnamefont {Holleis}},
  \bibinfo {author} {\bibfnamefont {Y.}~\bibnamefont {Saito}}, \bibinfo
  {author} {\bibfnamefont {L.}~\bibnamefont {Cohen}}, \bibinfo {author}
  {\bibfnamefont {W.}~\bibnamefont {Huynh}}, \bibinfo {author} {\bibfnamefont
  {C.~L.}\ \bibnamefont {Patterson}}, \bibinfo {author} {\bibfnamefont
  {F.}~\bibnamefont {Yang}}, \bibinfo {author} {\bibfnamefont {T.}~\bibnamefont
  {Taniguchi}}, \bibinfo {author} {\bibfnamefont {K.}~\bibnamefont {Watanabe}},
  \ and\ \bibinfo {author} {\bibfnamefont {A.~F.}\ \bibnamefont {Young}},\
  }\href {\doibase 10.1126/science.abm8386} {\bibfield  {journal} {\bibinfo
  {journal} {Science}\ }\textbf {\bibinfo {volume} {375}},\ \bibinfo {pages}
  {774} (\bibinfo {year} {2022})}\BibitemShut {NoStop}%
\bibitem [{\citenamefont {Zheng}\ \emph {et~al.}(2020)\citenamefont {Zheng},
  \citenamefont {Ma}, \citenamefont {Bi}, \citenamefont {de~la Barrera},
  \citenamefont {Liu}, \citenamefont {Mao}, \citenamefont {Zhang},
  \citenamefont {Kiper}, \citenamefont {Watanabe}, \citenamefont {Taniguchi},
  \citenamefont {Kong}, \citenamefont {Tisdale}, \citenamefont {Ashoori},
  \citenamefont {Gedik}, \citenamefont {Fu}, \citenamefont {Xu},\ and\
  \citenamefont {Jarillo-Herrero}}]{Zheng2020}%
  \BibitemOpen
  \bibfield  {author} {\bibinfo {author} {\bibfnamefont {Z.}~\bibnamefont
  {Zheng}}, \bibinfo {author} {\bibfnamefont {Q.}~\bibnamefont {Ma}}, \bibinfo
  {author} {\bibfnamefont {Z.}~\bibnamefont {Bi}}, \bibinfo {author}
  {\bibfnamefont {S.}~\bibnamefont {de~la Barrera}}, \bibinfo {author}
  {\bibfnamefont {M.-H.}\ \bibnamefont {Liu}}, \bibinfo {author} {\bibfnamefont
  {N.}~\bibnamefont {Mao}}, \bibinfo {author} {\bibfnamefont {Y.}~\bibnamefont
  {Zhang}}, \bibinfo {author} {\bibfnamefont {N.}~\bibnamefont {Kiper}},
  \bibinfo {author} {\bibfnamefont {K.}~\bibnamefont {Watanabe}}, \bibinfo
  {author} {\bibfnamefont {T.}~\bibnamefont {Taniguchi}}, \bibinfo {author}
  {\bibfnamefont {J.}~\bibnamefont {Kong}}, \bibinfo {author} {\bibfnamefont
  {W.~A.}\ \bibnamefont {Tisdale}}, \bibinfo {author} {\bibfnamefont
  {R.}~\bibnamefont {Ashoori}}, \bibinfo {author} {\bibfnamefont
  {N.}~\bibnamefont {Gedik}}, \bibinfo {author} {\bibfnamefont
  {L.}~\bibnamefont {Fu}}, \bibinfo {author} {\bibfnamefont {S.-Y.}\
  \bibnamefont {Xu}}, \ and\ \bibinfo {author} {\bibfnamefont {P.}~\bibnamefont
  {Jarillo-Herrero}},\ }\href {\doibase 10.1038/s41586-020-2970-9} {\bibfield
  {journal} {\bibinfo  {journal} {Nature}\ }\textbf {\bibinfo {volume} {588}},\
  \bibinfo {pages} {71} (\bibinfo {year} {2020})}\BibitemShut {NoStop}%
\bibitem [{\citenamefont {Overweg}\ \emph
  {et~al.}(2018{\natexlab{a}})\citenamefont {Overweg}, \citenamefont
  {Eggimann}, \citenamefont {Chen}, \citenamefont {Slizovskiy}, \citenamefont
  {Eich}, \citenamefont {Pisoni}, \citenamefont {Lee}, \citenamefont
  {Rickhaus}, \citenamefont {Watanabe}, \citenamefont {Taniguchi},
  \citenamefont {Fal'ko}, \citenamefont {Ihn},\ and\ \citenamefont
  {Ensslin}}]{Overweg2018}%
  \BibitemOpen
  \bibfield  {author} {\bibinfo {author} {\bibfnamefont {H.}~\bibnamefont
  {Overweg}}, \bibinfo {author} {\bibfnamefont {H.}~\bibnamefont {Eggimann}},
  \bibinfo {author} {\bibfnamefont {X.}~\bibnamefont {Chen}}, \bibinfo {author}
  {\bibfnamefont {S.}~\bibnamefont {Slizovskiy}}, \bibinfo {author}
  {\bibfnamefont {M.}~\bibnamefont {Eich}}, \bibinfo {author} {\bibfnamefont
  {R.}~\bibnamefont {Pisoni}}, \bibinfo {author} {\bibfnamefont
  {Y.}~\bibnamefont {Lee}}, \bibinfo {author} {\bibfnamefont {P.}~\bibnamefont
  {Rickhaus}}, \bibinfo {author} {\bibfnamefont {K.}~\bibnamefont {Watanabe}},
  \bibinfo {author} {\bibfnamefont {T.}~\bibnamefont {Taniguchi}}, \bibinfo
  {author} {\bibfnamefont {V.}~\bibnamefont {Fal'ko}}, \bibinfo {author}
  {\bibfnamefont {T.}~\bibnamefont {Ihn}}, \ and\ \bibinfo {author}
  {\bibfnamefont {K.}~\bibnamefont {Ensslin}},\ }\href {\doibase
  10.1021/acs.nanolett.7b04666} {\bibfield  {journal} {\bibinfo  {journal}
  {Nano Lett.}\ }\textbf {\bibinfo {volume} {18}},\ \bibinfo {pages} {553}
  (\bibinfo {year} {2018}{\natexlab{a}})}\BibitemShut {NoStop}%
\bibitem [{\citenamefont {Kraft}\ \emph {et~al.}(2018)\citenamefont {Kraft},
  \citenamefont {Krainov}, \citenamefont {Gall}, \citenamefont {Dmitriev},
  \citenamefont {Krupke}, \citenamefont {Gornyi},\ and\ \citenamefont
  {Danneau}}]{Kraft2018}%
  \BibitemOpen
  \bibfield  {author} {\bibinfo {author} {\bibfnamefont {R.}~\bibnamefont
  {Kraft}}, \bibinfo {author} {\bibfnamefont {I.~V.}\ \bibnamefont {Krainov}},
  \bibinfo {author} {\bibfnamefont {V.}~\bibnamefont {Gall}}, \bibinfo {author}
  {\bibfnamefont {A.~P.}\ \bibnamefont {Dmitriev}}, \bibinfo {author}
  {\bibfnamefont {R.}~\bibnamefont {Krupke}}, \bibinfo {author} {\bibfnamefont
  {I.~V.}\ \bibnamefont {Gornyi}}, \ and\ \bibinfo {author} {\bibfnamefont
  {R.}~\bibnamefont {Danneau}},\ }\href {\doibase
  10.1103/PhysRevLett.121.257703} {\bibfield  {journal} {\bibinfo  {journal}
  {Phys. Rev. Lett.}\ }\textbf {\bibinfo {volume} {121}},\ \bibinfo {pages}
  {257703} (\bibinfo {year} {2018})}\BibitemShut {NoStop}%
\bibitem [{\citenamefont {Overweg}\ \emph
  {et~al.}(2018{\natexlab{b}})\citenamefont {Overweg}, \citenamefont {Knothe},
  \citenamefont {Fabian}, \citenamefont {Linhart}, \citenamefont {Rickhaus},
  \citenamefont {Wernli}, \citenamefont {Watanabe}, \citenamefont {Taniguchi},
  \citenamefont {S\'anchez}, \citenamefont {Burgd\"orfer}, \citenamefont
  {Libisch}, \citenamefont {Fal'ko}, \citenamefont {Ensslin},\ and\
  \citenamefont {Ihn}}]{Overveg2018topo}%
  \BibitemOpen
  \bibfield  {author} {\bibinfo {author} {\bibfnamefont {H.}~\bibnamefont
  {Overweg}}, \bibinfo {author} {\bibfnamefont {A.}~\bibnamefont {Knothe}},
  \bibinfo {author} {\bibfnamefont {T.}~\bibnamefont {Fabian}}, \bibinfo
  {author} {\bibfnamefont {L.}~\bibnamefont {Linhart}}, \bibinfo {author}
  {\bibfnamefont {P.}~\bibnamefont {Rickhaus}}, \bibinfo {author}
  {\bibfnamefont {L.}~\bibnamefont {Wernli}}, \bibinfo {author} {\bibfnamefont
  {K.}~\bibnamefont {Watanabe}}, \bibinfo {author} {\bibfnamefont
  {T.}~\bibnamefont {Taniguchi}}, \bibinfo {author} {\bibfnamefont
  {D.}~\bibnamefont {S\'anchez}}, \bibinfo {author} {\bibfnamefont
  {J.}~\bibnamefont {Burgd\"orfer}}, \bibinfo {author} {\bibfnamefont
  {F.}~\bibnamefont {Libisch}}, \bibinfo {author} {\bibfnamefont {V.~I.}\
  \bibnamefont {Fal'ko}}, \bibinfo {author} {\bibfnamefont {K.}~\bibnamefont
  {Ensslin}}, \ and\ \bibinfo {author} {\bibfnamefont {T.}~\bibnamefont
  {Ihn}},\ }\href {\doibase 10.1103/PhysRevLett.121.257702} {\bibfield
  {journal} {\bibinfo  {journal} {Phys. Rev. Lett.}\ }\textbf {\bibinfo
  {volume} {121}},\ \bibinfo {pages} {257702} (\bibinfo {year}
  {2018}{\natexlab{b}})}\BibitemShut {NoStop}%
\bibitem [{\citenamefont {Banszerus}\ \emph
  {et~al.}(2020{\natexlab{a}})\citenamefont {Banszerus}, \citenamefont
  {M{\"o}ller}, \citenamefont {Icking}, \citenamefont {Watanabe}, \citenamefont
  {Taniguchi}, \citenamefont {Volk},\ and\ \citenamefont
  {Stampfer}}]{Banszerus2020}%
  \BibitemOpen
  \bibfield  {author} {\bibinfo {author} {\bibfnamefont {L.}~\bibnamefont
  {Banszerus}}, \bibinfo {author} {\bibfnamefont {S.}~\bibnamefont
  {M{\"o}ller}}, \bibinfo {author} {\bibfnamefont {E.}~\bibnamefont {Icking}},
  \bibinfo {author} {\bibfnamefont {K.}~\bibnamefont {Watanabe}}, \bibinfo
  {author} {\bibfnamefont {T.}~\bibnamefont {Taniguchi}}, \bibinfo {author}
  {\bibfnamefont {C.}~\bibnamefont {Volk}}, \ and\ \bibinfo {author}
  {\bibfnamefont {C.}~\bibnamefont {Stampfer}},\ }\href {\doibase
  10.1021/acs.nanolett.9b05295} {\bibfield  {journal} {\bibinfo  {journal}
  {Nano Lett.}\ }\textbf {\bibinfo {volume} {20}},\ \bibinfo {pages} {2005}
  (\bibinfo {year} {2020}{\natexlab{a}})}\BibitemShut {NoStop}%
\bibitem [{\citenamefont {Lee}\ \emph {et~al.}(2020)\citenamefont {Lee},
  \citenamefont {Knothe}, \citenamefont {Overweg}, \citenamefont {Eich},
  \citenamefont {Gold}, \citenamefont {Kurzmann}, \citenamefont {Klasovika},
  \citenamefont {Taniguchi}, \citenamefont {Wantanabe}, \citenamefont {Fal'ko},
  \citenamefont {Ihn}, \citenamefont {Ensslin},\ and\ \citenamefont
  {Rickhaus}}]{Lee2020}%
  \BibitemOpen
  \bibfield  {author} {\bibinfo {author} {\bibfnamefont {Y.}~\bibnamefont
  {Lee}}, \bibinfo {author} {\bibfnamefont {A.}~\bibnamefont {Knothe}},
  \bibinfo {author} {\bibfnamefont {H.}~\bibnamefont {Overweg}}, \bibinfo
  {author} {\bibfnamefont {M.}~\bibnamefont {Eich}}, \bibinfo {author}
  {\bibfnamefont {C.}~\bibnamefont {Gold}}, \bibinfo {author} {\bibfnamefont
  {A.}~\bibnamefont {Kurzmann}}, \bibinfo {author} {\bibfnamefont
  {V.}~\bibnamefont {Klasovika}}, \bibinfo {author} {\bibfnamefont
  {T.}~\bibnamefont {Taniguchi}}, \bibinfo {author} {\bibfnamefont
  {K.}~\bibnamefont {Wantanabe}}, \bibinfo {author} {\bibfnamefont
  {V.}~\bibnamefont {Fal'ko}}, \bibinfo {author} {\bibfnamefont
  {T.}~\bibnamefont {Ihn}}, \bibinfo {author} {\bibfnamefont {K.}~\bibnamefont
  {Ensslin}}, \ and\ \bibinfo {author} {\bibfnamefont {P.}~\bibnamefont
  {Rickhaus}},\ }\href {\doibase 10.1103/PhysRevLett.124.126802} {\bibfield
  {journal} {\bibinfo  {journal} {Phys. Rev. Lett.}\ }\textbf {\bibinfo
  {volume} {124}},\ \bibinfo {pages} {126802} (\bibinfo {year}
  {2020})}\BibitemShut {NoStop}%
\bibitem [{\citenamefont {Allen}\ \emph {et~al.}(2012)\citenamefont {Allen},
  \citenamefont {Martin},\ and\ \citenamefont {Yacoby}}]{Allen2012}%
  \BibitemOpen
  \bibfield  {author} {\bibinfo {author} {\bibfnamefont {M.~T.}\ \bibnamefont
  {Allen}}, \bibinfo {author} {\bibfnamefont {J.}~\bibnamefont {Martin}}, \
  and\ \bibinfo {author} {\bibfnamefont {A.}~\bibnamefont {Yacoby}},\ }\href
  {\doibase 10.1038/ncomms1945} {\bibfield  {journal} {\bibinfo  {journal}
  {Nat. Commun.}\ }\textbf {\bibinfo {volume} {3}},\ \bibinfo {pages} {934}
  (\bibinfo {year} {2012})}\BibitemShut {NoStop}%
\bibitem [{\citenamefont {Goossens}\ \emph {et~al.}(2012)\citenamefont
  {Goossens}, \citenamefont {Driessen}, \citenamefont {Baart}, \citenamefont
  {Watanabe}, \citenamefont {Taniguchi},\ and\ \citenamefont
  {Vandersypen}}]{Goossens2012}%
  \BibitemOpen
  \bibfield  {author} {\bibinfo {author} {\bibfnamefont {A.~S.~M.}\
  \bibnamefont {Goossens}}, \bibinfo {author} {\bibfnamefont {S.~C.~M.}\
  \bibnamefont {Driessen}}, \bibinfo {author} {\bibfnamefont {T.~A.}\
  \bibnamefont {Baart}}, \bibinfo {author} {\bibfnamefont {K.}~\bibnamefont
  {Watanabe}}, \bibinfo {author} {\bibfnamefont {T.}~\bibnamefont {Taniguchi}},
  \ and\ \bibinfo {author} {\bibfnamefont {L.~M.~K.}\ \bibnamefont
  {Vandersypen}},\ }\href {\doibase 10.1021/nl301986q} {\bibfield  {journal}
  {\bibinfo  {journal} {Nano Lett.}\ }\textbf {\bibinfo {volume} {12}},\
  \bibinfo {pages} {4656} (\bibinfo {year} {2012})}\BibitemShut {NoStop}%
\bibitem [{\citenamefont {Eich}\ \emph {et~al.}(2018)\citenamefont {Eich},
  \citenamefont {Herman}, \citenamefont {Pisoni}, \citenamefont {Overweg},
  \citenamefont {Kurzmann}, \citenamefont {Lee}, \citenamefont {Rickhaus},
  \citenamefont {Watanabe}, \citenamefont {Taniguchi}, \citenamefont {Sigrist},
  \citenamefont {Ihn},\ and\ \citenamefont {Ensslin}}]{Eich2018}%
  \BibitemOpen
  \bibfield  {author} {\bibinfo {author} {\bibfnamefont {M.}~\bibnamefont
  {Eich}}, \bibinfo {author} {\bibfnamefont {F.~c.~v.}\ \bibnamefont {Herman}},
  \bibinfo {author} {\bibfnamefont {R.}~\bibnamefont {Pisoni}}, \bibinfo
  {author} {\bibfnamefont {H.}~\bibnamefont {Overweg}}, \bibinfo {author}
  {\bibfnamefont {A.}~\bibnamefont {Kurzmann}}, \bibinfo {author}
  {\bibfnamefont {Y.}~\bibnamefont {Lee}}, \bibinfo {author} {\bibfnamefont
  {P.}~\bibnamefont {Rickhaus}}, \bibinfo {author} {\bibfnamefont
  {K.}~\bibnamefont {Watanabe}}, \bibinfo {author} {\bibfnamefont
  {T.}~\bibnamefont {Taniguchi}}, \bibinfo {author} {\bibfnamefont
  {M.}~\bibnamefont {Sigrist}}, \bibinfo {author} {\bibfnamefont
  {T.}~\bibnamefont {Ihn}}, \ and\ \bibinfo {author} {\bibfnamefont
  {K.}~\bibnamefont {Ensslin}},\ }\href {\doibase 10.1103/PhysRevX.8.031023}
  {\bibfield  {journal} {\bibinfo  {journal} {Phys. Rev. X}\ }\textbf {\bibinfo
  {volume} {8}},\ \bibinfo {pages} {031023} (\bibinfo {year}
  {2018})}\BibitemShut {NoStop}%
\bibitem [{\citenamefont {Kurzmann}\ \emph {et~al.}(2019)\citenamefont
  {Kurzmann}, \citenamefont {Overweg}, \citenamefont {Eich}, \citenamefont
  {Pally}, \citenamefont {Rickhaus}, \citenamefont {Pisoni}, \citenamefont
  {Lee}, \citenamefont {Watanabe}, \citenamefont {Taniguchi}, \citenamefont
  {Ihn},\ and\ \citenamefont {Ensslin}}]{Kurzmann2019}%
  \BibitemOpen
  \bibfield  {author} {\bibinfo {author} {\bibfnamefont {A.}~\bibnamefont
  {Kurzmann}}, \bibinfo {author} {\bibfnamefont {H.}~\bibnamefont {Overweg}},
  \bibinfo {author} {\bibfnamefont {M.}~\bibnamefont {Eich}}, \bibinfo {author}
  {\bibfnamefont {A.}~\bibnamefont {Pally}}, \bibinfo {author} {\bibfnamefont
  {P.}~\bibnamefont {Rickhaus}}, \bibinfo {author} {\bibfnamefont
  {R.}~\bibnamefont {Pisoni}}, \bibinfo {author} {\bibfnamefont
  {Y.}~\bibnamefont {Lee}}, \bibinfo {author} {\bibfnamefont {K.}~\bibnamefont
  {Watanabe}}, \bibinfo {author} {\bibfnamefont {T.}~\bibnamefont {Taniguchi}},
  \bibinfo {author} {\bibfnamefont {T.}~\bibnamefont {Ihn}}, \ and\ \bibinfo
  {author} {\bibfnamefont {K.}~\bibnamefont {Ensslin}},\ }\href {\doibase
  10.1021/acs.nanolett.9b01617} {\bibfield  {journal} {\bibinfo  {journal}
  {Nano Lett.}\ }\textbf {\bibinfo {volume} {19}},\ \bibinfo {pages} {5216}
  (\bibinfo {year} {2019})}\BibitemShut {NoStop}%
\bibitem [{\citenamefont {Seemann}\ \emph {et~al.}(2023)\citenamefont
  {Seemann}, \citenamefont {Knothe},\ and\ \citenamefont
  {Hentschel}}]{Seemann2023}%
  \BibitemOpen
  \bibfield  {author} {\bibinfo {author} {\bibfnamefont {L.}~\bibnamefont
  {Seemann}}, \bibinfo {author} {\bibfnamefont {A.}~\bibnamefont {Knothe}}, \
  and\ \bibinfo {author} {\bibfnamefont {M.}~\bibnamefont {Hentschel}},\ }\href
  {\doibase 10.1103/PhysRevB.107.205404} {\bibfield  {journal} {\bibinfo
  {journal} {Phys. Rev. B}\ }\textbf {\bibinfo {volume} {107}},\ \bibinfo
  {pages} {205404} (\bibinfo {year} {2023})}\BibitemShut {NoStop}%
\bibitem [{\citenamefont {Martin}\ \emph {et~al.}(2008)\citenamefont {Martin},
  \citenamefont {Blanter},\ and\ \citenamefont {Morpurgo}}]{Martin2008}%
  \BibitemOpen
  \bibfield  {author} {\bibinfo {author} {\bibfnamefont {I.}~\bibnamefont
  {Martin}}, \bibinfo {author} {\bibfnamefont {Y.~M.}\ \bibnamefont {Blanter}},
  \ and\ \bibinfo {author} {\bibfnamefont {A.~F.}\ \bibnamefont {Morpurgo}},\
  }\href {\doibase 10.1103/PhysRevLett.100.036804} {\bibfield  {journal}
  {\bibinfo  {journal} {Phys. Rev. Lett.}\ }\textbf {\bibinfo {volume} {100}},\
  \bibinfo {pages} {036804} (\bibinfo {year} {2008})}\BibitemShut {NoStop}%
\bibitem [{\citenamefont {San-Jose}\ \emph {et~al.}(2009)\citenamefont
  {San-Jose}, \citenamefont {Prada}, \citenamefont {McCann},\ and\
  \citenamefont {Schomerus}}]{San-Jose2009}%
  \BibitemOpen
  \bibfield  {author} {\bibinfo {author} {\bibfnamefont {P.}~\bibnamefont
  {San-Jose}}, \bibinfo {author} {\bibfnamefont {E.}~\bibnamefont {Prada}},
  \bibinfo {author} {\bibfnamefont {E.}~\bibnamefont {McCann}}, \ and\ \bibinfo
  {author} {\bibfnamefont {H.}~\bibnamefont {Schomerus}},\ }\href {\doibase
  10.1103/PhysRevLett.102.247204} {\bibfield  {journal} {\bibinfo  {journal}
  {Phys. Rev. Lett.}\ }\textbf {\bibinfo {volume} {102}},\ \bibinfo {pages}
  {247204} (\bibinfo {year} {2009})}\BibitemShut {NoStop}%
\bibitem [{\citenamefont {Li}\ \emph {et~al.}(2016)\citenamefont {Li},
  \citenamefont {Wang}, \citenamefont {McFaul}, \citenamefont {Zern},
  \citenamefont {Ren}, \citenamefont {Watanabe}, \citenamefont {Taniguchi},
  \citenamefont {Qiao},\ and\ \citenamefont {Zhu}}]{Li2016}%
  \BibitemOpen
  \bibfield  {author} {\bibinfo {author} {\bibfnamefont {J.}~\bibnamefont
  {Li}}, \bibinfo {author} {\bibfnamefont {K.}~\bibnamefont {Wang}}, \bibinfo
  {author} {\bibfnamefont {K.~J.}\ \bibnamefont {McFaul}}, \bibinfo {author}
  {\bibfnamefont {Z.}~\bibnamefont {Zern}}, \bibinfo {author} {\bibfnamefont
  {Y.}~\bibnamefont {Ren}}, \bibinfo {author} {\bibfnamefont {K.}~\bibnamefont
  {Watanabe}}, \bibinfo {author} {\bibfnamefont {T.}~\bibnamefont {Taniguchi}},
  \bibinfo {author} {\bibfnamefont {Z.}~\bibnamefont {Qiao}}, \ and\ \bibinfo
  {author} {\bibfnamefont {J.}~\bibnamefont {Zhu}},\ }\href {\doibase
  10.1038/nnano.2016.158} {\bibfield  {journal} {\bibinfo  {journal} {Nat.
  Nanotechnol.}\ }\textbf {\bibinfo {volume} {11}},\ \bibinfo {pages} {1060}
  (\bibinfo {year} {2016})}\BibitemShut {NoStop}%
\bibitem [{\citenamefont {Ingla-Ayn{\'e}s}\ \emph {et~al.}(2023)\citenamefont
  {Ingla-Ayn{\'e}s}, \citenamefont {Manesco}, \citenamefont {Ghiasi},
  \citenamefont {Volosheniuk}, \citenamefont {Watanabe}, \citenamefont
  {Taniguchi},\ and\ \citenamefont {van~der Zant}}]{Ingla-Aynes2023}%
  \BibitemOpen
  \bibfield  {author} {\bibinfo {author} {\bibfnamefont {J.}~\bibnamefont
  {Ingla-Ayn{\'e}s}}, \bibinfo {author} {\bibfnamefont {A.~L.~R.}\ \bibnamefont
  {Manesco}}, \bibinfo {author} {\bibfnamefont {T.~S.}\ \bibnamefont {Ghiasi}},
  \bibinfo {author} {\bibfnamefont {S.}~\bibnamefont {Volosheniuk}}, \bibinfo
  {author} {\bibfnamefont {K.}~\bibnamefont {Watanabe}}, \bibinfo {author}
  {\bibfnamefont {T.}~\bibnamefont {Taniguchi}}, \ and\ \bibinfo {author}
  {\bibfnamefont {H.~S.~J.}\ \bibnamefont {van~der Zant}},\ }\href {\doibase
  10.1021/acs.nanolett.3c00499} {\bibfield  {journal} {\bibinfo  {journal}
  {Nano Lett.}\ }\textbf {\bibinfo {volume} {23}},\ \bibinfo {pages} {5453}
  (\bibinfo {year} {2023})}\BibitemShut {NoStop}%
\bibitem [{\citenamefont {Iwakiri}\ \emph {et~al.}(2022)\citenamefont
  {Iwakiri}, \citenamefont {de~Vries}, \citenamefont {Portol{\'e}s},
  \citenamefont {Zheng}, \citenamefont {Taniguchi}, \citenamefont {Watanabe},
  \citenamefont {Ihn},\ and\ \citenamefont {Ensslin}}]{Iwakiri2022}%
  \BibitemOpen
  \bibfield  {author} {\bibinfo {author} {\bibfnamefont {S.}~\bibnamefont
  {Iwakiri}}, \bibinfo {author} {\bibfnamefont {F.~K.}\ \bibnamefont
  {de~Vries}}, \bibinfo {author} {\bibfnamefont {E.}~\bibnamefont
  {Portol{\'e}s}}, \bibinfo {author} {\bibfnamefont {G.}~\bibnamefont {Zheng}},
  \bibinfo {author} {\bibfnamefont {T.}~\bibnamefont {Taniguchi}}, \bibinfo
  {author} {\bibfnamefont {K.}~\bibnamefont {Watanabe}}, \bibinfo {author}
  {\bibfnamefont {T.}~\bibnamefont {Ihn}}, \ and\ \bibinfo {author}
  {\bibfnamefont {K.}~\bibnamefont {Ensslin}},\ }\href {\doibase
  10.1021/acs.nanolett.2c01874} {\bibfield  {journal} {\bibinfo  {journal}
  {Nano Lett.}\ }\textbf {\bibinfo {volume} {22}},\ \bibinfo {pages} {6292}
  (\bibinfo {year} {2022})}\BibitemShut {NoStop}%
\bibitem [{\citenamefont {Fu}\ \emph {et~al.}(2023)\citenamefont {Fu},
  \citenamefont {Huang}, \citenamefont {Watanabe}, \citenamefont {Taniguchi},
  \citenamefont {Kayyalha},\ and\ \citenamefont {Zhu}}]{Fu2023}%
  \BibitemOpen
  \bibfield  {author} {\bibinfo {author} {\bibfnamefont {H.}~\bibnamefont
  {Fu}}, \bibinfo {author} {\bibfnamefont {K.}~\bibnamefont {Huang}}, \bibinfo
  {author} {\bibfnamefont {K.}~\bibnamefont {Watanabe}}, \bibinfo {author}
  {\bibfnamefont {T.}~\bibnamefont {Taniguchi}}, \bibinfo {author}
  {\bibfnamefont {M.}~\bibnamefont {Kayyalha}}, \ and\ \bibinfo {author}
  {\bibfnamefont {J.}~\bibnamefont {Zhu}},\ }\href {\doibase
  10.1021/acs.nanolett.2c05004} {\bibfield  {journal} {\bibinfo  {journal}
  {Nano Lett.}\ }\textbf {\bibinfo {volume} {23}},\ \bibinfo {pages} {718}
  (\bibinfo {year} {2023})}\BibitemShut {NoStop}%
\bibitem [{\citenamefont {Varlet}\ \emph {et~al.}(2014)\citenamefont {Varlet},
  \citenamefont {Liu}, \citenamefont {Krueckl}, \citenamefont {Bischoff},
  \citenamefont {Simonet}, \citenamefont {Watanabe}, \citenamefont {Taniguchi},
  \citenamefont {Richter}, \citenamefont {Ensslin},\ and\ \citenamefont
  {Ihn}}]{Varlet2014}%
  \BibitemOpen
  \bibfield  {author} {\bibinfo {author} {\bibfnamefont {A.}~\bibnamefont
  {Varlet}}, \bibinfo {author} {\bibfnamefont {M.-H.}\ \bibnamefont {Liu}},
  \bibinfo {author} {\bibfnamefont {V.}~\bibnamefont {Krueckl}}, \bibinfo
  {author} {\bibfnamefont {D.}~\bibnamefont {Bischoff}}, \bibinfo {author}
  {\bibfnamefont {P.}~\bibnamefont {Simonet}}, \bibinfo {author} {\bibfnamefont
  {K.}~\bibnamefont {Watanabe}}, \bibinfo {author} {\bibfnamefont
  {T.}~\bibnamefont {Taniguchi}}, \bibinfo {author} {\bibfnamefont
  {K.}~\bibnamefont {Richter}}, \bibinfo {author} {\bibfnamefont
  {K.}~\bibnamefont {Ensslin}}, \ and\ \bibinfo {author} {\bibfnamefont
  {T.}~\bibnamefont {Ihn}},\ }\href {\doibase 10.1103/PhysRevLett.113.116601}
  {\bibfield  {journal} {\bibinfo  {journal} {Phys. Rev. Lett.}\ }\textbf
  {\bibinfo {volume} {113}},\ \bibinfo {pages} {116601} (\bibinfo {year}
  {2014})}\BibitemShut {NoStop}%
\bibitem [{\citenamefont {Du}\ \emph {et~al.}(2018)\citenamefont {Du},
  \citenamefont {Liu}, \citenamefont {Mohrmann}, \citenamefont {Wu},
  \citenamefont {Krupke}, \citenamefont {von L\"ohneysen}, \citenamefont
  {Richter},\ and\ \citenamefont {Danneau}}]{Du2018}%
  \BibitemOpen
  \bibfield  {author} {\bibinfo {author} {\bibfnamefont {R.}~\bibnamefont
  {Du}}, \bibinfo {author} {\bibfnamefont {M.-H.}\ \bibnamefont {Liu}},
  \bibinfo {author} {\bibfnamefont {J.}~\bibnamefont {Mohrmann}}, \bibinfo
  {author} {\bibfnamefont {F.}~\bibnamefont {Wu}}, \bibinfo {author}
  {\bibfnamefont {R.}~\bibnamefont {Krupke}}, \bibinfo {author} {\bibfnamefont
  {H.}~\bibnamefont {von L\"ohneysen}}, \bibinfo {author} {\bibfnamefont
  {K.}~\bibnamefont {Richter}}, \ and\ \bibinfo {author} {\bibfnamefont
  {R.}~\bibnamefont {Danneau}},\ }\href {\doibase
  10.1103/PhysRevLett.121.127706} {\bibfield  {journal} {\bibinfo  {journal}
  {Phys. Rev. Lett.}\ }\textbf {\bibinfo {volume} {121}},\ \bibinfo {pages}
  {127706} (\bibinfo {year} {2018})}\BibitemShut {NoStop}%
\bibitem [{\citenamefont {Liu}\ \emph {et~al.}(2015)\citenamefont {Liu},
  \citenamefont {Rickhaus}, \citenamefont {Makk}, \citenamefont {T\'ov\'ari},
  \citenamefont {Maurand}, \citenamefont {Tkatschenko}, \citenamefont {Weiss},
  \citenamefont {Sch\"onenberger},\ and\ \citenamefont
  {Richter}}]{Liu2015scalable}%
  \BibitemOpen
  \bibfield  {author} {\bibinfo {author} {\bibfnamefont {M.-H.}\ \bibnamefont
  {Liu}}, \bibinfo {author} {\bibfnamefont {P.}~\bibnamefont {Rickhaus}},
  \bibinfo {author} {\bibfnamefont {P.}~\bibnamefont {Makk}}, \bibinfo {author}
  {\bibfnamefont {E.}~\bibnamefont {T\'ov\'ari}}, \bibinfo {author}
  {\bibfnamefont {R.}~\bibnamefont {Maurand}}, \bibinfo {author} {\bibfnamefont
  {F.}~\bibnamefont {Tkatschenko}}, \bibinfo {author} {\bibfnamefont
  {M.}~\bibnamefont {Weiss}}, \bibinfo {author} {\bibfnamefont
  {C.}~\bibnamefont {Sch\"onenberger}}, \ and\ \bibinfo {author} {\bibfnamefont
  {K.}~\bibnamefont {Richter}},\ }\href {\doibase
  10.1103/PhysRevLett.114.036601} {\bibfield  {journal} {\bibinfo  {journal}
  {Phys. Rev. Lett.}\ }\textbf {\bibinfo {volume} {114}},\ \bibinfo {pages}
  {036601} (\bibinfo {year} {2015})}\BibitemShut {NoStop}%
\bibitem [{\citenamefont {Kaplan}(1992)}]{Kaplan1992}%
  \BibitemOpen
  \bibfield  {author} {\bibinfo {author} {\bibfnamefont {D.~B.}\ \bibnamefont
  {Kaplan}},\ }\href {\doibase https://doi.org/10.1016/0370-2693(92)91112-M}
  {\bibfield  {journal} {\bibinfo  {journal} {Phys. Lett. B}\ }\textbf
  {\bibinfo {volume} {288}},\ \bibinfo {pages} {342} (\bibinfo {year}
  {1992})}\BibitemShut {NoStop}%
\bibitem [{\citenamefont {Drell}\ \emph {et~al.}(1976)\citenamefont {Drell},
  \citenamefont {Weinstein},\ and\ \citenamefont {Yankielowicz}}]{Drell1976}%
  \BibitemOpen
  \bibfield  {author} {\bibinfo {author} {\bibfnamefont {S.~D.}\ \bibnamefont
  {Drell}}, \bibinfo {author} {\bibfnamefont {M.}~\bibnamefont {Weinstein}}, \
  and\ \bibinfo {author} {\bibfnamefont {S.}~\bibnamefont {Yankielowicz}},\
  }\href {\doibase 10.1103/PhysRevD.14.1627} {\bibfield  {journal} {\bibinfo
  {journal} {Phys. Rev. D}\ }\textbf {\bibinfo {volume} {14}},\ \bibinfo
  {pages} {1627} (\bibinfo {year} {1976})}\BibitemShut {NoStop}%
\bibitem [{\citenamefont {Wilson}(1975)}]{Wilson1977}%
  \BibitemOpen
  \bibfield  {author} {\bibinfo {author} {\bibfnamefont {K.~G.}\ \bibnamefont
  {Wilson}},\ }in\ \href@noop {} {\emph {\bibinfo {booktitle} {{13th
  International School of Subnuclear Physics: New Phenomena in Subnuclear
  Physics}}}}\ (\bibinfo {year} {1975})\BibitemShut {NoStop}%
\bibitem [{\citenamefont {Susskind}(1977)}]{Susskind1977}%
  \BibitemOpen
  \bibfield  {author} {\bibinfo {author} {\bibfnamefont {L.}~\bibnamefont
  {Susskind}},\ }\href {\doibase 10.1103/PhysRevD.16.3031} {\bibfield
  {journal} {\bibinfo  {journal} {Phys. Rev. D}\ }\textbf {\bibinfo {volume}
  {16}},\ \bibinfo {pages} {3031} (\bibinfo {year} {1977})}\BibitemShut
  {NoStop}%
\bibitem [{\citenamefont {Stacey}(1982)}]{Stacey1982}%
  \BibitemOpen
  \bibfield  {author} {\bibinfo {author} {\bibfnamefont {R.}~\bibnamefont
  {Stacey}},\ }\href {\doibase 10.1103/PhysRevD.26.468} {\bibfield  {journal}
  {\bibinfo  {journal} {Phys. Rev. D}\ }\textbf {\bibinfo {volume} {26}},\
  \bibinfo {pages} {468} (\bibinfo {year} {1982})}\BibitemShut {NoStop}%
\bibitem [{\citenamefont {Kogut}(1983)}]{Kogut1983}%
  \BibitemOpen
  \bibfield  {author} {\bibinfo {author} {\bibfnamefont {J.~B.}\ \bibnamefont
  {Kogut}},\ }\href {\doibase 10.1103/RevModPhys.55.775} {\bibfield  {journal}
  {\bibinfo  {journal} {Rev. Mod. Phys.}\ }\textbf {\bibinfo {volume} {55}},\
  \bibinfo {pages} {775} (\bibinfo {year} {1983})}\BibitemShut {NoStop}%
\bibitem [{\citenamefont {Beenakker}\ \emph {et~al.}(2023)\citenamefont
  {Beenakker}, \citenamefont {Don\'is~Vela}, \citenamefont {Lemut},
  \citenamefont {Pacholski},\ and\ \citenamefont
  {Tworzyd\l{}o}}]{Beenakker2023}%
  \BibitemOpen
  \bibfield  {author} {\bibinfo {author} {\bibfnamefont {C.~W.~J.}\
  \bibnamefont {Beenakker}}, \bibinfo {author} {\bibfnamefont {A.}~\bibnamefont
  {Don\'is~Vela}}, \bibinfo {author} {\bibfnamefont {G.}~\bibnamefont {Lemut}},
  \bibinfo {author} {\bibfnamefont {M.~J.}\ \bibnamefont {Pacholski}}, \ and\
  \bibinfo {author} {\bibfnamefont {J.}~\bibnamefont {Tworzyd\l{}o}},\ }\href
  {\doibase https://doi.org/10.1002/andp.202300081} {\bibfield  {journal}
  {\bibinfo  {journal} {Ann. Phys.}\ }\textbf {\bibinfo {volume} {535}},\
  \bibinfo {pages} {2300081} (\bibinfo {year} {2023})},\ \Eprint
  {http://arxiv.org/abs/https://onlinelibrary.wiley.com/doi/pdf/10.1002/andp.202300081}
  {https://onlinelibrary.wiley.com/doi/pdf/10.1002/andp.202300081} \BibitemShut
  {NoStop}%
\bibitem [{\citenamefont {Tworzyd\l{}o}\ \emph {et~al.}(2008)\citenamefont
  {Tworzyd\l{}o}, \citenamefont {Groth},\ and\ \citenamefont
  {Beenakker}}]{Tworzydlo2008}%
  \BibitemOpen
  \bibfield  {author} {\bibinfo {author} {\bibfnamefont {J.}~\bibnamefont
  {Tworzyd\l{}o}}, \bibinfo {author} {\bibfnamefont {C.~W.}\ \bibnamefont
  {Groth}}, \ and\ \bibinfo {author} {\bibfnamefont {C.~W.~J.}\ \bibnamefont
  {Beenakker}},\ }\href {\doibase 10.1103/PhysRevB.78.235438} {\bibfield
  {journal} {\bibinfo  {journal} {Phys. Rev. B}\ }\textbf {\bibinfo {volume}
  {78}},\ \bibinfo {pages} {235438} (\bibinfo {year} {2008})}\BibitemShut
  {NoStop}%
\bibitem [{\citenamefont {Hern\'andez}\ and\ \citenamefont
  {Lewenkopf}(2012)}]{Hernandez2012}%
  \BibitemOpen
  \bibfield  {author} {\bibinfo {author} {\bibfnamefont {A.~R.}\ \bibnamefont
  {Hern\'andez}}\ and\ \bibinfo {author} {\bibfnamefont {C.~H.}\ \bibnamefont
  {Lewenkopf}},\ }\href {\doibase 10.1103/PhysRevB.86.155439} {\bibfield
  {journal} {\bibinfo  {journal} {Phys. Rev. B}\ }\textbf {\bibinfo {volume}
  {86}},\ \bibinfo {pages} {155439} (\bibinfo {year} {2012})}\BibitemShut
  {NoStop}%
\bibitem [{\citenamefont {Habib}\ \emph {et~al.}(2016)\citenamefont {Habib},
  \citenamefont {Sajjad},\ and\ \citenamefont {Ghosh}}]{Habib2016}%
  \BibitemOpen
  \bibfield  {author} {\bibinfo {author} {\bibfnamefont {K.~M.~M.}\
  \bibnamefont {Habib}}, \bibinfo {author} {\bibfnamefont {R.~N.}\ \bibnamefont
  {Sajjad}}, \ and\ \bibinfo {author} {\bibfnamefont {A.~W.}\ \bibnamefont
  {Ghosh}},\ }\href {\doibase 10.1063/1.4943790} {\bibfield  {journal}
  {\bibinfo  {journal} {Appl. Phys. Lett.}\ }\textbf {\bibinfo {volume}
  {108}},\ \bibinfo {pages} {113105} (\bibinfo {year} {2016})},\ \Eprint
  {http://arxiv.org/abs/https://pubs.aip.org/aip/apl/article-pdf/doi/10.1063/1.4943790/14478056/113105\_1\_online.pdf}
  {https://pubs.aip.org/aip/apl/article-pdf/doi/10.1063/1.4943790/14478056/113105\_1\_online.pdf}
  \BibitemShut {NoStop}%
\bibitem [{\citenamefont {Szafran}\ \emph {et~al.}(2019)\citenamefont
  {Szafran}, \citenamefont {Mre\ifmmode \acute{n}\else
  \'{n}\fi{}ca-Kolasi\ifmmode~\acute{n}\else \'{n}\fi{}ska},\ and\
  \citenamefont {\ifmmode~\dot{Z}\else \.{Z}\fi{}ebrowski}}]{Szafran2019}%
  \BibitemOpen
  \bibfield  {author} {\bibinfo {author} {\bibfnamefont {B.}~\bibnamefont
  {Szafran}}, \bibinfo {author} {\bibfnamefont {A.}~\bibnamefont {Mre\ifmmode
  \acute{n}\else \'{n}\fi{}ca-Kolasi\ifmmode~\acute{n}\else \'{n}\fi{}ska}}, \
  and\ \bibinfo {author} {\bibfnamefont {D.}~\bibnamefont
  {\ifmmode~\dot{Z}\else \.{Z}\fi{}ebrowski}},\ }\href {\doibase
  10.1103/PhysRevB.99.195406} {\bibfield  {journal} {\bibinfo  {journal} {Phys.
  Rev. B}\ }\textbf {\bibinfo {volume} {99}},\ \bibinfo {pages} {195406}
  (\bibinfo {year} {2019})}\BibitemShut {NoStop}%
\bibitem [{\citenamefont {Ziesen}\ \emph {et~al.}(2023)\citenamefont {Ziesen},
  \citenamefont {Fulga},\ and\ \citenamefont {Hassler}}]{Ziesen2023}%
  \BibitemOpen
  \bibfield  {author} {\bibinfo {author} {\bibfnamefont {A.}~\bibnamefont
  {Ziesen}}, \bibinfo {author} {\bibfnamefont {I.~C.}\ \bibnamefont {Fulga}}, \
  and\ \bibinfo {author} {\bibfnamefont {F.}~\bibnamefont {Hassler}},\ }\href
  {\doibase 10.1103/PhysRevB.107.195409} {\bibfield  {journal} {\bibinfo
  {journal} {Phys. Rev. B}\ }\textbf {\bibinfo {volume} {107}},\ \bibinfo
  {pages} {195409} (\bibinfo {year} {2023})}\BibitemShut {NoStop}%
\bibitem [{\citenamefont {\ifmmode~\dot{Z}\else \.{Z}\fi{}ebrowski}\ \emph
  {et~al.}(2017)\citenamefont {\ifmmode~\dot{Z}\else \.{Z}\fi{}ebrowski},
  \citenamefont {Peeters},\ and\ \citenamefont {Szafran}}]{Zebrowski2017}%
  \BibitemOpen
  \bibfield  {author} {\bibinfo {author} {\bibfnamefont {D.~P.}\ \bibnamefont
  {\ifmmode~\dot{Z}\else \.{Z}\fi{}ebrowski}}, \bibinfo {author} {\bibfnamefont
  {F.~M.}\ \bibnamefont {Peeters}}, \ and\ \bibinfo {author} {\bibfnamefont
  {B.}~\bibnamefont {Szafran}},\ }\href {\doibase 10.1103/PhysRevB.96.035434}
  {\bibfield  {journal} {\bibinfo  {journal} {Phys. Rev. B}\ }\textbf {\bibinfo
  {volume} {96}},\ \bibinfo {pages} {035434} (\bibinfo {year}
  {2017})}\BibitemShut {NoStop}%
\bibitem [{\citenamefont {Novoselov}\ \emph {et~al.}(2005)\citenamefont
  {Novoselov}, \citenamefont {Geim}, \citenamefont {Morozov}, \citenamefont
  {Jiang}, \citenamefont {Katsnelson}, \citenamefont {Grigorieva},
  \citenamefont {Dubonos},\ and\ \citenamefont {Firsov}}]{Novoselov2005}%
  \BibitemOpen
  \bibfield  {author} {\bibinfo {author} {\bibfnamefont {K.~S.}\ \bibnamefont
  {Novoselov}}, \bibinfo {author} {\bibfnamefont {A.~K.}\ \bibnamefont {Geim}},
  \bibinfo {author} {\bibfnamefont {S.~V.}\ \bibnamefont {Morozov}}, \bibinfo
  {author} {\bibfnamefont {D.}~\bibnamefont {Jiang}}, \bibinfo {author}
  {\bibfnamefont {M.~I.}\ \bibnamefont {Katsnelson}}, \bibinfo {author}
  {\bibfnamefont {I.~V.}\ \bibnamefont {Grigorieva}}, \bibinfo {author}
  {\bibfnamefont {S.~V.}\ \bibnamefont {Dubonos}}, \ and\ \bibinfo {author}
  {\bibfnamefont {A.~A.}\ \bibnamefont {Firsov}},\ }\href {\doibase
  10.1038/nature04233} {\bibfield  {journal} {\bibinfo  {journal} {Nature}\
  }\textbf {\bibinfo {volume} {438}},\ \bibinfo {pages} {197} (\bibinfo {year}
  {2005})}\BibitemShut {NoStop}%
\bibitem [{\citenamefont {Zhang}\ \emph {et~al.}(2005)\citenamefont {Zhang},
  \citenamefont {Tan}, \citenamefont {Stormer},\ and\ \citenamefont
  {Kim}}]{Zhang2005}%
  \BibitemOpen
  \bibfield  {author} {\bibinfo {author} {\bibfnamefont {Y.}~\bibnamefont
  {Zhang}}, \bibinfo {author} {\bibfnamefont {Y.-W.}\ \bibnamefont {Tan}},
  \bibinfo {author} {\bibfnamefont {H.~L.}\ \bibnamefont {Stormer}}, \ and\
  \bibinfo {author} {\bibfnamefont {P.}~\bibnamefont {Kim}},\ }\href {\doibase
  10.1038/nature04235} {\bibfield  {journal} {\bibinfo  {journal} {Nature}\
  }\textbf {\bibinfo {volume} {438}},\ \bibinfo {pages} {201} (\bibinfo {year}
  {2005})}\BibitemShut {NoStop}%
\bibitem [{\citenamefont {Taychatanapat}\ \emph {et~al.}(2013)\citenamefont
  {Taychatanapat}, \citenamefont {Watanabe}, \citenamefont {Taniguchi},\ and\
  \citenamefont {Jarillo-Herrero}}]{Taychatanapat2013}%
  \BibitemOpen
  \bibfield  {author} {\bibinfo {author} {\bibfnamefont {T.}~\bibnamefont
  {Taychatanapat}}, \bibinfo {author} {\bibfnamefont {K.}~\bibnamefont
  {Watanabe}}, \bibinfo {author} {\bibfnamefont {T.}~\bibnamefont {Taniguchi}},
  \ and\ \bibinfo {author} {\bibfnamefont {P.}~\bibnamefont
  {Jarillo-Herrero}},\ }\href {\doibase 10.1038/nphys2549} {\bibfield
  {journal} {\bibinfo  {journal} {Nat. Phys.}\ }\textbf {\bibinfo {volume}
  {9}},\ \bibinfo {pages} {225} (\bibinfo {year} {2013})}\BibitemShut {NoStop}%
\bibitem [{\citenamefont {McCann}\ and\ \citenamefont
  {Koshino}(2013)}]{McCann2013}%
  \BibitemOpen
  \bibfield  {author} {\bibinfo {author} {\bibfnamefont {E.}~\bibnamefont
  {McCann}}\ and\ \bibinfo {author} {\bibfnamefont {M.}~\bibnamefont
  {Koshino}},\ }\href {\doibase 10.1088/0034-4885/76/5/056503} {\bibfield
  {journal} {\bibinfo  {journal} {Rep. Progr. Phys.}\ }\textbf {\bibinfo
  {volume} {76}},\ \bibinfo {pages} {056503} (\bibinfo {year}
  {2013})}\BibitemShut {NoStop}%
\bibitem [{\citenamefont {Groth}\ \emph {et~al.}(2014)\citenamefont {Groth},
  \citenamefont {Wimmer}, \citenamefont {Akhmerov},\ and\ \citenamefont
  {Waintal}}]{Groth2014}%
  \BibitemOpen
  \bibfield  {author} {\bibinfo {author} {\bibfnamefont {C.~W.}\ \bibnamefont
  {Groth}}, \bibinfo {author} {\bibfnamefont {M.}~\bibnamefont {Wimmer}},
  \bibinfo {author} {\bibfnamefont {A.~R.}\ \bibnamefont {Akhmerov}}, \ and\
  \bibinfo {author} {\bibfnamefont {X.}~\bibnamefont {Waintal}},\ }\href
  {\doibase 10.1088/1367-2630/16/6/063065} {\bibfield  {journal} {\bibinfo
  {journal} {New. J. Phys.}\ }\textbf {\bibinfo {volume} {16}},\ \bibinfo
  {pages} {063065} (\bibinfo {year} {2014})}\BibitemShut {NoStop}%
\bibitem [{\citenamefont {Kolasi\ifmmode~\acute{n}\else \'{n}\fi{}ski}\ \emph
  {et~al.}(2016)\citenamefont {Kolasi\ifmmode~\acute{n}\else \'{n}\fi{}ski},
  \citenamefont {Szafran}, \citenamefont {Brun},\ and\ \citenamefont
  {Sellier}}]{Kol2016transport}%
  \BibitemOpen
  \bibfield  {author} {\bibinfo {author} {\bibfnamefont {K.}~\bibnamefont
  {Kolasi\ifmmode~\acute{n}\else \'{n}\fi{}ski}}, \bibinfo {author}
  {\bibfnamefont {B.}~\bibnamefont {Szafran}}, \bibinfo {author} {\bibfnamefont
  {B.}~\bibnamefont {Brun}}, \ and\ \bibinfo {author} {\bibfnamefont
  {H.}~\bibnamefont {Sellier}},\ }\href {\doibase 10.1103/PhysRevB.94.075301}
  {\bibfield  {journal} {\bibinfo  {journal} {Phys. Rev. B}\ }\textbf {\bibinfo
  {volume} {94}},\ \bibinfo {pages} {075301} (\bibinfo {year}
  {2016})}\BibitemShut {NoStop}%
\bibitem [{\citenamefont {Mre{\'{n}}ca-Kolasi{\'{n}}ska}\ \emph
  {et~al.}(2023)\citenamefont {Mre{\'{n}}ca-Kolasi{\'{n}}ska}, \citenamefont
  {Chen},\ and\ \citenamefont {Liu}}]{Mrenca2023}%
  \BibitemOpen
  \bibfield  {author} {\bibinfo {author} {\bibfnamefont {A.}~\bibnamefont
  {Mre{\'{n}}ca-Kolasi{\'{n}}ska}}, \bibinfo {author} {\bibfnamefont {S.-C.}\
  \bibnamefont {Chen}}, \ and\ \bibinfo {author} {\bibfnamefont {M.-H.}\
  \bibnamefont {Liu}},\ }\href {\doibase 10.1038/s41699-023-00426-9} {\bibfield
   {journal} {\bibinfo  {journal} {npj 2D Materials and Applications}\ }\textbf
  {\bibinfo {volume} {7}},\ \bibinfo {pages} {64} (\bibinfo {year}
  {2023})}\BibitemShut {NoStop}%
\bibitem [{\citenamefont {Aharonov}\ and\ \citenamefont
  {Bohm}(1959)}]{Aharonov1959}%
  \BibitemOpen
  \bibfield  {author} {\bibinfo {author} {\bibfnamefont {Y.}~\bibnamefont
  {Aharonov}}\ and\ \bibinfo {author} {\bibfnamefont {D.}~\bibnamefont
  {Bohm}},\ }\href {\doibase 10.1103/PhysRev.115.485} {\bibfield  {journal}
  {\bibinfo  {journal} {Phys. Rev.}\ }\textbf {\bibinfo {volume} {115}},\
  \bibinfo {pages} {485} (\bibinfo {year} {1959})}\BibitemShut {NoStop}%
\bibitem [{\citenamefont {Recher}\ \emph {et~al.}(2007)\citenamefont {Recher},
  \citenamefont {Trauzettel}, \citenamefont {Rycerz}, \citenamefont {Blanter},
  \citenamefont {Beenakker},\ and\ \citenamefont {Morpurgo}}]{Recher2007}%
  \BibitemOpen
  \bibfield  {author} {\bibinfo {author} {\bibfnamefont {P.}~\bibnamefont
  {Recher}}, \bibinfo {author} {\bibfnamefont {B.}~\bibnamefont {Trauzettel}},
  \bibinfo {author} {\bibfnamefont {A.}~\bibnamefont {Rycerz}}, \bibinfo
  {author} {\bibfnamefont {Y.~M.}\ \bibnamefont {Blanter}}, \bibinfo {author}
  {\bibfnamefont {C.~W.~J.}\ \bibnamefont {Beenakker}}, \ and\ \bibinfo
  {author} {\bibfnamefont {A.~F.}\ \bibnamefont {Morpurgo}},\ }\href {\doibase
  10.1103/PhysRevB.76.235404} {\bibfield  {journal} {\bibinfo  {journal} {Phys.
  Rev. B}\ }\textbf {\bibinfo {volume} {76}},\ \bibinfo {pages} {235404}
  (\bibinfo {year} {2007})}\BibitemShut {NoStop}%
\bibitem [{\citenamefont {Wurm}\ \emph {et~al.}(2010)\citenamefont {Wurm},
  \citenamefont {Wimmer}, \citenamefont {Baranger},\ and\ \citenamefont
  {Richter}}]{Wurm2010}%
  \BibitemOpen
  \bibfield  {author} {\bibinfo {author} {\bibfnamefont {J.}~\bibnamefont
  {Wurm}}, \bibinfo {author} {\bibfnamefont {M.}~\bibnamefont {Wimmer}},
  \bibinfo {author} {\bibfnamefont {H.~U.}\ \bibnamefont {Baranger}}, \ and\
  \bibinfo {author} {\bibfnamefont {K.}~\bibnamefont {Richter}},\ }\href
  {\doibase 10.1088/0268-1242/25/3/034003} {\bibfield  {journal} {\bibinfo
  {journal} {Semicond. Sci. Tech.}\ }\textbf {\bibinfo {volume} {25}},\
  \bibinfo {pages} {034003} (\bibinfo {year} {2010})}\BibitemShut {NoStop}%
\bibitem [{\citenamefont {Mre\ifmmode \acute{n}\else
  \'{n}\fi{}ca-Kolasi\ifmmode~\acute{n}\else \'{n}\fi{}ska}\ and\ \citenamefont
  {Szafran}(2016)}]{Mrenca2016}%
  \BibitemOpen
  \bibfield  {author} {\bibinfo {author} {\bibfnamefont {A.}~\bibnamefont
  {Mre\ifmmode \acute{n}\else \'{n}\fi{}ca-Kolasi\ifmmode~\acute{n}\else
  \'{n}\fi{}ska}}\ and\ \bibinfo {author} {\bibfnamefont {B.}~\bibnamefont
  {Szafran}},\ }\href {\doibase 10.1103/PhysRevB.94.195315} {\bibfield
  {journal} {\bibinfo  {journal} {Phys. Rev. B}\ }\textbf {\bibinfo {volume}
  {94}},\ \bibinfo {pages} {195315} (\bibinfo {year} {2016})}\BibitemShut
  {NoStop}%
\bibitem [{\citenamefont {Russo}\ \emph {et~al.}(2008)\citenamefont {Russo},
  \citenamefont {Oostinga}, \citenamefont {Wehenkel}, \citenamefont {Heersche},
  \citenamefont {Sobhani}, \citenamefont {Vandersypen},\ and\ \citenamefont
  {Morpurgo}}]{Russo2008}%
  \BibitemOpen
  \bibfield  {author} {\bibinfo {author} {\bibfnamefont {S.}~\bibnamefont
  {Russo}}, \bibinfo {author} {\bibfnamefont {J.~B.}\ \bibnamefont {Oostinga}},
  \bibinfo {author} {\bibfnamefont {D.}~\bibnamefont {Wehenkel}}, \bibinfo
  {author} {\bibfnamefont {H.~B.}\ \bibnamefont {Heersche}}, \bibinfo {author}
  {\bibfnamefont {S.~S.}\ \bibnamefont {Sobhani}}, \bibinfo {author}
  {\bibfnamefont {L.~M.~K.}\ \bibnamefont {Vandersypen}}, \ and\ \bibinfo
  {author} {\bibfnamefont {A.~F.}\ \bibnamefont {Morpurgo}},\ }\href {\doibase
  10.1103/PhysRevB.77.085413} {\bibfield  {journal} {\bibinfo  {journal} {Phys.
  Rev. B}\ }\textbf {\bibinfo {volume} {77}},\ \bibinfo {pages} {085413}
  (\bibinfo {year} {2008})}\BibitemShut {NoStop}%
\bibitem [{\citenamefont {Huefner}\ \emph {et~al.}(2010)\citenamefont
  {Huefner}, \citenamefont {Molitor}, \citenamefont {Jacobsen}, \citenamefont
  {Pioda}, \citenamefont {Stampfer}, \citenamefont {Ensslin},\ and\
  \citenamefont {Ihn}}]{Huefner2010}%
  \BibitemOpen
  \bibfield  {author} {\bibinfo {author} {\bibfnamefont {M.}~\bibnamefont
  {Huefner}}, \bibinfo {author} {\bibfnamefont {F.}~\bibnamefont {Molitor}},
  \bibinfo {author} {\bibfnamefont {A.}~\bibnamefont {Jacobsen}}, \bibinfo
  {author} {\bibfnamefont {A.}~\bibnamefont {Pioda}}, \bibinfo {author}
  {\bibfnamefont {C.}~\bibnamefont {Stampfer}}, \bibinfo {author}
  {\bibfnamefont {K.}~\bibnamefont {Ensslin}}, \ and\ \bibinfo {author}
  {\bibfnamefont {T.}~\bibnamefont {Ihn}},\ }\href {\doibase
  10.1088/1367-2630/12/4/043054} {\bibfield  {journal} {\bibinfo  {journal}
  {New. J. Phys.}\ }\textbf {\bibinfo {volume} {12}},\ \bibinfo {pages}
  {043054} (\bibinfo {year} {2010})}\BibitemShut {NoStop}%
\bibitem [{\citenamefont {Smirnov}\ \emph {et~al.}(2012)\citenamefont
  {Smirnov}, \citenamefont {Schmidt},\ and\ \citenamefont
  {Haug}}]{Smirnov2012}%
  \BibitemOpen
  \bibfield  {author} {\bibinfo {author} {\bibfnamefont {D.}~\bibnamefont
  {Smirnov}}, \bibinfo {author} {\bibfnamefont {H.}~\bibnamefont {Schmidt}}, \
  and\ \bibinfo {author} {\bibfnamefont {R.~J.}\ \bibnamefont {Haug}},\ }\href
  {\doibase 10.1063/1.4717622} {\bibfield  {journal} {\bibinfo  {journal}
  {Appl. Phys. Lett.}\ }\textbf {\bibinfo {volume} {100}},\ \bibinfo {pages}
  {203114} (\bibinfo {year} {2012})}\BibitemShut {NoStop}%
\bibitem [{\citenamefont {Smirnov}\ \emph {et~al.}(2014)\citenamefont
  {Smirnov}, \citenamefont {Rode},\ and\ \citenamefont {Haug}}]{Smirnov2014}%
  \BibitemOpen
  \bibfield  {author} {\bibinfo {author} {\bibfnamefont {D.}~\bibnamefont
  {Smirnov}}, \bibinfo {author} {\bibfnamefont {J.~C.}\ \bibnamefont {Rode}}, \
  and\ \bibinfo {author} {\bibfnamefont {R.~J.}\ \bibnamefont {Haug}},\ }\href
  {\doibase 10.1063/1.4894471} {\bibfield  {journal} {\bibinfo  {journal}
  {Appl. Phys. Lett.}\ }\textbf {\bibinfo {volume} {105}},\ \bibinfo {pages}
  {082112} (\bibinfo {year} {2014})}\BibitemShut {NoStop}%
\bibitem [{\citenamefont {Dauber}\ \emph {et~al.}(2017)\citenamefont {Dauber},
  \citenamefont {Oellers}, \citenamefont {Venn}, \citenamefont {Epping},
  \citenamefont {Watanabe}, \citenamefont {Taniguchi}, \citenamefont
  {Hassler},\ and\ \citenamefont {Stampfer}}]{Dauber2017}%
  \BibitemOpen
  \bibfield  {author} {\bibinfo {author} {\bibfnamefont {J.}~\bibnamefont
  {Dauber}}, \bibinfo {author} {\bibfnamefont {M.}~\bibnamefont {Oellers}},
  \bibinfo {author} {\bibfnamefont {F.}~\bibnamefont {Venn}}, \bibinfo {author}
  {\bibfnamefont {A.}~\bibnamefont {Epping}}, \bibinfo {author} {\bibfnamefont
  {K.}~\bibnamefont {Watanabe}}, \bibinfo {author} {\bibfnamefont
  {T.}~\bibnamefont {Taniguchi}}, \bibinfo {author} {\bibfnamefont
  {F.}~\bibnamefont {Hassler}}, \ and\ \bibinfo {author} {\bibfnamefont
  {C.}~\bibnamefont {Stampfer}},\ }\href {\doibase 10.1103/PhysRevB.96.205407}
  {\bibfield  {journal} {\bibinfo  {journal} {Phys. Rev. B}\ }\textbf {\bibinfo
  {volume} {96}},\ \bibinfo {pages} {205407} (\bibinfo {year}
  {2017})}\BibitemShut {NoStop}%
\bibitem [{\citenamefont {Ronen}\ \emph {et~al.}(2021)\citenamefont {Ronen},
  \citenamefont {Werkmeister}, \citenamefont {Haie~Najafabadi}, \citenamefont
  {Pierce}, \citenamefont {Anderson}, \citenamefont {Shin}, \citenamefont
  {Lee}, \citenamefont {Lee}, \citenamefont {Johnson}, \citenamefont
  {Watanabe}, \citenamefont {Taniguchi}, \citenamefont {Yacoby},\ and\
  \citenamefont {Kim}}]{Ronen2021}%
  \BibitemOpen
  \bibfield  {author} {\bibinfo {author} {\bibfnamefont {Y.}~\bibnamefont
  {Ronen}}, \bibinfo {author} {\bibfnamefont {T.}~\bibnamefont {Werkmeister}},
  \bibinfo {author} {\bibfnamefont {D.}~\bibnamefont {Haie~Najafabadi}},
  \bibinfo {author} {\bibfnamefont {A.~T.}\ \bibnamefont {Pierce}}, \bibinfo
  {author} {\bibfnamefont {L.~E.}\ \bibnamefont {Anderson}}, \bibinfo {author}
  {\bibfnamefont {Y.~J.}\ \bibnamefont {Shin}}, \bibinfo {author}
  {\bibfnamefont {S.~Y.}\ \bibnamefont {Lee}}, \bibinfo {author} {\bibfnamefont
  {Y.~H.}\ \bibnamefont {Lee}}, \bibinfo {author} {\bibfnamefont
  {B.}~\bibnamefont {Johnson}}, \bibinfo {author} {\bibfnamefont
  {K.}~\bibnamefont {Watanabe}}, \bibinfo {author} {\bibfnamefont
  {T.}~\bibnamefont {Taniguchi}}, \bibinfo {author} {\bibfnamefont
  {A.}~\bibnamefont {Yacoby}}, \ and\ \bibinfo {author} {\bibfnamefont
  {P.}~\bibnamefont {Kim}},\ }\href {\doibase 10.1038/s41565-021-00861-z}
  {\bibfield  {journal} {\bibinfo  {journal} {Nat. Nanotechnol.}\ }\textbf
  {\bibinfo {volume} {16}},\ \bibinfo {pages} {563} (\bibinfo {year}
  {2021})}\BibitemShut {NoStop}%
\bibitem [{\citenamefont {D{\'e}prez}\ \emph {et~al.}(2021)\citenamefont
  {D{\'e}prez}, \citenamefont {Veyrat}, \citenamefont {Vignaud}, \citenamefont
  {Nayak}, \citenamefont {Watanabe}, \citenamefont {Taniguchi}, \citenamefont
  {Gay}, \citenamefont {Sellier},\ and\ \citenamefont
  {Sac{\'e}p{\'e}}}]{Deprez2021}%
  \BibitemOpen
  \bibfield  {author} {\bibinfo {author} {\bibfnamefont {C.}~\bibnamefont
  {D{\'e}prez}}, \bibinfo {author} {\bibfnamefont {L.}~\bibnamefont {Veyrat}},
  \bibinfo {author} {\bibfnamefont {H.}~\bibnamefont {Vignaud}}, \bibinfo
  {author} {\bibfnamefont {G.}~\bibnamefont {Nayak}}, \bibinfo {author}
  {\bibfnamefont {K.}~\bibnamefont {Watanabe}}, \bibinfo {author}
  {\bibfnamefont {T.}~\bibnamefont {Taniguchi}}, \bibinfo {author}
  {\bibfnamefont {F.}~\bibnamefont {Gay}}, \bibinfo {author} {\bibfnamefont
  {H.}~\bibnamefont {Sellier}}, \ and\ \bibinfo {author} {\bibfnamefont
  {B.}~\bibnamefont {Sac{\'e}p{\'e}}},\ }\href {\doibase
  10.1038/s41565-021-00847-x} {\bibfield  {journal} {\bibinfo  {journal} {Nat.
  Nanotechnol.}\ }\textbf {\bibinfo {volume} {16}},\ \bibinfo {pages} {555}
  (\bibinfo {year} {2021})}\BibitemShut {NoStop}%
\bibitem [{\citenamefont {Iwakiri}\ \emph {et~al.}(2023)\citenamefont
  {Iwakiri}, \citenamefont {Mestre-Tor\'a}, \citenamefont {Portol\'es},
  \citenamefont {Visscher}, \citenamefont {Perego}, \citenamefont {Zheng},
  \citenamefont {Taniguchi}, \citenamefont {Watanabe}, \citenamefont {Sigrist},
  \citenamefont {Ihn},\ and\ \citenamefont {Ensslin}}]{Iwakiri2023}%
  \BibitemOpen
  \bibfield  {author} {\bibinfo {author} {\bibfnamefont {S.}~\bibnamefont
  {Iwakiri}}, \bibinfo {author} {\bibfnamefont {A.}~\bibnamefont
  {Mestre-Tor\'a}}, \bibinfo {author} {\bibfnamefont {E.}~\bibnamefont
  {Portol\'es}}, \bibinfo {author} {\bibfnamefont {M.}~\bibnamefont
  {Visscher}}, \bibinfo {author} {\bibfnamefont {M.}~\bibnamefont {Perego}},
  \bibinfo {author} {\bibfnamefont {G.}~\bibnamefont {Zheng}}, \bibinfo
  {author} {\bibfnamefont {T.}~\bibnamefont {Taniguchi}}, \bibinfo {author}
  {\bibfnamefont {K.}~\bibnamefont {Watanabe}}, \bibinfo {author}
  {\bibfnamefont {M.}~\bibnamefont {Sigrist}}, \bibinfo {author} {\bibfnamefont
  {T.}~\bibnamefont {Ihn}}, \ and\ \bibinfo {author} {\bibfnamefont
  {K.}~\bibnamefont {Ensslin}},\ }\href@noop {} {\enquote {\bibinfo {title}
  {Tunable quantum interferometer for correlated moir\'e electrons},}\ }
  (\bibinfo {year} {2023}),\ \Eprint {http://arxiv.org/abs/2308.07400}
  {arXiv:2308.07400 [cond-mat.mes-hall]} \BibitemShut {NoStop}%
\bibitem [{\citenamefont {Hansen}\ \emph {et~al.}(2001)\citenamefont {Hansen},
  \citenamefont {Kristensen}, \citenamefont {Pedersen}, \citenamefont
  {S\o{}rensen},\ and\ \citenamefont {Lindelof}}]{Hansen2001}%
  \BibitemOpen
  \bibfield  {author} {\bibinfo {author} {\bibfnamefont {A.~E.}\ \bibnamefont
  {Hansen}}, \bibinfo {author} {\bibfnamefont {A.}~\bibnamefont {Kristensen}},
  \bibinfo {author} {\bibfnamefont {S.}~\bibnamefont {Pedersen}}, \bibinfo
  {author} {\bibfnamefont {C.~B.}\ \bibnamefont {S\o{}rensen}}, \ and\ \bibinfo
  {author} {\bibfnamefont {P.~E.}\ \bibnamefont {Lindelof}},\ }\href {\doibase
  10.1103/PhysRevB.64.045327} {\bibfield  {journal} {\bibinfo  {journal} {Phys.
  Rev. B}\ }\textbf {\bibinfo {volume} {64}},\ \bibinfo {pages} {045327}
  (\bibinfo {year} {2001})}\BibitemShut {NoStop}%
\bibitem [{\citenamefont {Wimmer}(2008)}]{Wimmer2008}%
  \BibitemOpen
  \bibfield  {author} {\bibinfo {author} {\bibfnamefont {M.}~\bibnamefont
  {Wimmer}},\ }\emph {\bibinfo {title} {Quantum transport in nanostructures:
  From computational concepts to spintronics in graphene and magnetictunnel
  junctions}},\ \href@noop {} {Ph.D. thesis},\ \bibinfo  {school} {Universitat
  Regensburg} (\bibinfo {year} {2008})\BibitemShut {NoStop}%
\bibitem [{\citenamefont {Liu}\ and\ \citenamefont
  {Richter}(2012)}]{2012periodicModel}%
  \BibitemOpen
  \bibfield  {author} {\bibinfo {author} {\bibfnamefont {M.-H.}\ \bibnamefont
  {Liu}}\ and\ \bibinfo {author} {\bibfnamefont {K.}~\bibnamefont {Richter}},\
  }\href {\doibase 10.1103/PhysRevB.86.115455} {\bibfield  {journal} {\bibinfo
  {journal} {Phys. Rev. B}\ }\textbf {\bibinfo {volume} {86}},\ \bibinfo
  {pages} {115455} (\bibinfo {year} {2012})}\BibitemShut {NoStop}%
\bibitem [{\citenamefont {Liu}\ \emph {et~al.}(2012)\citenamefont {Liu},
  \citenamefont {Bundesmann},\ and\ \citenamefont {Richter}}]{Liu2012periodic}%
  \BibitemOpen
  \bibfield  {author} {\bibinfo {author} {\bibfnamefont {M.-H.}\ \bibnamefont
  {Liu}}, \bibinfo {author} {\bibfnamefont {J.}~\bibnamefont {Bundesmann}}, \
  and\ \bibinfo {author} {\bibfnamefont {K.}~\bibnamefont {Richter}},\ }\href
  {\doibase 10.1103/PhysRevB.85.085406} {\bibfield  {journal} {\bibinfo
  {journal} {Phys. Rev. B}\ }\textbf {\bibinfo {volume} {85}},\ \bibinfo
  {pages} {085406} (\bibinfo {year} {2012})}\BibitemShut {NoStop}%
\bibitem [{\citenamefont {Ghorashi}\ \emph {et~al.}(2023)\citenamefont
  {Ghorashi}, \citenamefont {Dunbrack}, \citenamefont {Abouelkomsan},
  \citenamefont {Sun}, \citenamefont {Du},\ and\ \citenamefont
  {Cano}}]{Ghorashi2023}%
  \BibitemOpen
  \bibfield  {author} {\bibinfo {author} {\bibfnamefont {S.~A.~A.}\
  \bibnamefont {Ghorashi}}, \bibinfo {author} {\bibfnamefont {A.}~\bibnamefont
  {Dunbrack}}, \bibinfo {author} {\bibfnamefont {A.}~\bibnamefont
  {Abouelkomsan}}, \bibinfo {author} {\bibfnamefont {J.}~\bibnamefont {Sun}},
  \bibinfo {author} {\bibfnamefont {X.}~\bibnamefont {Du}}, \ and\ \bibinfo
  {author} {\bibfnamefont {J.}~\bibnamefont {Cano}},\ }\href {\doibase
  10.1103/PhysRevLett.130.196201} {\bibfield  {journal} {\bibinfo  {journal}
  {Phys. Rev. Lett.}\ }\textbf {\bibinfo {volume} {130}},\ \bibinfo {pages}
  {196201} (\bibinfo {year} {2023})}\BibitemShut {NoStop}%
\bibitem [{\citenamefont {Sun}\ \emph {et~al.}(2023)\citenamefont {Sun},
  \citenamefont {Ghorashi}, \citenamefont {Watanabe}, \citenamefont
  {Taniguchi}, \citenamefont {Camino}, \citenamefont {Cano},\ and\
  \citenamefont {Du}}]{Sun2023}%
  \BibitemOpen
  \bibfield  {author} {\bibinfo {author} {\bibfnamefont {J.}~\bibnamefont
  {Sun}}, \bibinfo {author} {\bibfnamefont {S.~A.~A.}\ \bibnamefont
  {Ghorashi}}, \bibinfo {author} {\bibfnamefont {K.}~\bibnamefont {Watanabe}},
  \bibinfo {author} {\bibfnamefont {T.}~\bibnamefont {Taniguchi}}, \bibinfo
  {author} {\bibfnamefont {F.}~\bibnamefont {Camino}}, \bibinfo {author}
  {\bibfnamefont {J.}~\bibnamefont {Cano}}, \ and\ \bibinfo {author}
  {\bibfnamefont {X.}~\bibnamefont {Du}},\ }\href@noop {} {\enquote {\bibinfo
  {title} {{Signature of Correlated Insulator in Electric Field Controlled
  Superlattice}},}\ } (\bibinfo {year} {2023}),\ \Eprint
  {http://arxiv.org/abs/2306.06848} {arXiv:2306.06848 [cond-mat.str-el]}
  \BibitemShut {NoStop}%
\bibitem [{\citenamefont {Banszerus}\ \emph {et~al.}(2018)\citenamefont
  {Banszerus}, \citenamefont {Frohn}, \citenamefont {Epping}, \citenamefont
  {Neumaier}, \citenamefont {Watanabe}, \citenamefont {Taniguchi},\ and\
  \citenamefont {Stampfer}}]{Banszerus2018}%
  \BibitemOpen
  \bibfield  {author} {\bibinfo {author} {\bibfnamefont {L.}~\bibnamefont
  {Banszerus}}, \bibinfo {author} {\bibfnamefont {B.}~\bibnamefont {Frohn}},
  \bibinfo {author} {\bibfnamefont {A.}~\bibnamefont {Epping}}, \bibinfo
  {author} {\bibfnamefont {D.}~\bibnamefont {Neumaier}}, \bibinfo {author}
  {\bibfnamefont {K.}~\bibnamefont {Watanabe}}, \bibinfo {author}
  {\bibfnamefont {T.}~\bibnamefont {Taniguchi}}, \ and\ \bibinfo {author}
  {\bibfnamefont {C.}~\bibnamefont {Stampfer}},\ }\href {\doibase
  10.1021/acs.nanolett.8b01303} {\bibfield  {journal} {\bibinfo  {journal}
  {Nano Lett.}\ }\textbf {\bibinfo {volume} {18}},\ \bibinfo {pages} {4785}
  (\bibinfo {year} {2018})}\BibitemShut {NoStop}%
\bibitem [{\citenamefont {Banszerus}\ \emph
  {et~al.}(2020{\natexlab{b}})\citenamefont {Banszerus}, \citenamefont {Frohn},
  \citenamefont {Fabian}, \citenamefont {Somanchi}, \citenamefont {Epping},
  \citenamefont {M\"uller}, \citenamefont {Neumaier}, \citenamefont {Watanabe},
  \citenamefont {Taniguchi}, \citenamefont {Libisch}, \citenamefont
  {Beschoten}, \citenamefont {Hassler},\ and\ \citenamefont
  {Stampfer}}]{Banszerus2020dbl}%
  \BibitemOpen
  \bibfield  {author} {\bibinfo {author} {\bibfnamefont {L.}~\bibnamefont
  {Banszerus}}, \bibinfo {author} {\bibfnamefont {B.}~\bibnamefont {Frohn}},
  \bibinfo {author} {\bibfnamefont {T.}~\bibnamefont {Fabian}}, \bibinfo
  {author} {\bibfnamefont {S.}~\bibnamefont {Somanchi}}, \bibinfo {author}
  {\bibfnamefont {A.}~\bibnamefont {Epping}}, \bibinfo {author} {\bibfnamefont
  {M.}~\bibnamefont {M\"uller}}, \bibinfo {author} {\bibfnamefont
  {D.}~\bibnamefont {Neumaier}}, \bibinfo {author} {\bibfnamefont
  {K.}~\bibnamefont {Watanabe}}, \bibinfo {author} {\bibfnamefont
  {T.}~\bibnamefont {Taniguchi}}, \bibinfo {author} {\bibfnamefont
  {F.}~\bibnamefont {Libisch}}, \bibinfo {author} {\bibfnamefont
  {B.}~\bibnamefont {Beschoten}}, \bibinfo {author} {\bibfnamefont
  {F.}~\bibnamefont {Hassler}}, \ and\ \bibinfo {author} {\bibfnamefont
  {C.}~\bibnamefont {Stampfer}},\ }\href {\doibase
  10.1103/PhysRevLett.124.177701} {\bibfield  {journal} {\bibinfo  {journal}
  {Phys. Rev. Lett.}\ }\textbf {\bibinfo {volume} {124}},\ \bibinfo {pages}
  {177701} (\bibinfo {year} {2020}{\natexlab{b}})}\BibitemShut {NoStop}%
\bibitem [{\citenamefont {Gold}\ \emph {et~al.}(2021)\citenamefont {Gold},
  \citenamefont {Knothe}, \citenamefont {Kurzmann}, \citenamefont
  {Garcia-Ruiz}, \citenamefont {Watanabe}, \citenamefont {Taniguchi},
  \citenamefont {Fal'ko}, \citenamefont {Ensslin},\ and\ \citenamefont
  {Ihn}}]{Gold2021}%
  \BibitemOpen
  \bibfield  {author} {\bibinfo {author} {\bibfnamefont {C.}~\bibnamefont
  {Gold}}, \bibinfo {author} {\bibfnamefont {A.}~\bibnamefont {Knothe}},
  \bibinfo {author} {\bibfnamefont {A.}~\bibnamefont {Kurzmann}}, \bibinfo
  {author} {\bibfnamefont {A.}~\bibnamefont {Garcia-Ruiz}}, \bibinfo {author}
  {\bibfnamefont {K.}~\bibnamefont {Watanabe}}, \bibinfo {author}
  {\bibfnamefont {T.}~\bibnamefont {Taniguchi}}, \bibinfo {author}
  {\bibfnamefont {V.}~\bibnamefont {Fal'ko}}, \bibinfo {author} {\bibfnamefont
  {K.}~\bibnamefont {Ensslin}}, \ and\ \bibinfo {author} {\bibfnamefont
  {T.}~\bibnamefont {Ihn}},\ }\href {\doibase 10.1103/PhysRevLett.127.046801}
  {\bibfield  {journal} {\bibinfo  {journal} {Phys. Rev. Lett.}\ }\textbf
  {\bibinfo {volume} {127}},\ \bibinfo {pages} {046801} (\bibinfo {year}
  {2021})}\BibitemShut {NoStop}%
\end{thebibliography}%

% \appendix

\renewcommand{\thefigure}{S\arabic{figure}}
\renewcommand{\theequation}{S\arabic{equation}}
\renewcommand{\thesection}{S\arabic{section}}

\setcounter{figure}{0}
\setcounter{equation}{0}

\clearpage
\part{Supplemental Material}

\section{Effective model band structure}
It is instructive to derive the energy bands within the effective model, and see its low-energy properties. For an artificial infinite 2D square lattice, the Hamiltonian can be expressed as
\begin{equation}
H_{s} = \left(
\begin{matrix}
	U/2 	&	h_{12}	&	\gamma_1	&	0	\\
	h_{21}	&	U/2		&	0			&	0	\\
\gamma_1	&	0		&	- U/2		&	h_{34} \\
	0		&	0		&	h_{43}	&	- U/2	\\
\end{matrix}
\right),
\label{eq:H_eff_2d}
\end{equation}
where 
\begin{equation}
\begin{aligned}
h_{12} = h^*_{21} &= \frac{it}{2}(-e^{ik_xa} + e^{-ik_xa}) + \frac{t}{2}(e^{ik_ya} - e^{-ik_ya}) \\
&= t (\sin k_x a + i \sin k_y a),
\label{eq:h12}
\end{aligned}
\end{equation}
and similarly $h_{34} = h^*_{43} = t (\sin k_x a - i \sin k_y a)$. 
Diagonalizing (\autoref{eq:H_eff_2d}), we obtain \autoref{eq:E_eff_2d} in the main text.

\section{Low-$k$ expansion of the band structure}
As described in the main text, in the effective model, the number of the $K$ and $K'$ valleys is doubled [\autoref{fig:fig1}(d), upper panel]. 
However, out of the doubly degenerate cones, those in the middle of the edge of the Brillouin zone, $\mathbf{k}=(\pm \pi/a, 0)$ and $\mathbf{k}=(0, \pm \pi/a)$, exhibit the electronic properties of the $K'$ valley of BLG, while those in the corners of the Brillouin zone -- of the $K$ valley. 
This can be seen for example by substituting $\mathbf{k}=(k_x+ \pi/a, k_y)$ in (\autoref{eq:H_eff_2d}). Then, the matrix elements become 
\begin{equation}
\begin{aligned}
h_{12} = h^*_{21} = t (\sin k_x a - i \sin k_y a) \approx t(k_xa - ik_ya), \\
h_{34} = h^*_{43} = t (\sin k_x a + i \sin k_y a) \approx t(k_xa + ik_ya), 
\end{aligned}
\end{equation} 
where the last step is the approximate form at low energy. With $t=\hbar v_f/a$, the Hamiltonian matrix is same as (\autoref{eq:H_blg}), with $\xi=-1$ and $V=0$. 
By analogy, at $\mathbf{k}=(k_x+ \pi/a, k_y+ \pi/a)$, one can arrive at $h_{12} = h^*_{21} = -t (k_x a + i k_y a)$ and $h_{34} = h^*_{43} = -t (k_x a - i k_y a)$, which corresponds to the $K$ valley Hamiltonian.

\section{Band offset and asymmetry parameter}
\label{app:band_offset}
Considering the parallel-plate capacitor model, we obtain the carrier density $n(x,y)$ in the device with (\autoref{eq:dens}). For a given value of $n$ at the point $(x,y)$, we calculate the asymmetry parameter $U$ as described in the Supplemental Material of Ref.\ \cite{Varlet2014}.

Given the carrier density $n$ and $U$, one can obtain the band offset from (\autoref{eq:E_eff_2d}). For this, we notice that the energy dispersion is to a good approximation isotropic at $|E|\lesssim 1$ eV, %[see \autoref{fig:figS1}(b)],
an energy range well suited for quantum the transport calculations. 
(\autoref{eq:E_eff_2d}) can be simplified by considering the direction along $k_y=0$, and substituting $(\sin^2k_xa + \sin^2k_ya)$ for $\sin^2k_xa$. Then, the asymmetry parameter is obtained by replacing $k_x$ by $\sqrt{\pi|n|}$ and adding a minus sign
\begin{equation}
\begin{aligned}
V &= -\sgn(n) \left( 
t^2 \sin^2 (\sqrt{\pi n}a) + \frac{U^2}{4} + \frac{\gamma_1^2}{2} \right. \\
&\left. - \frac{1}{2}
\left[ 
\gamma_1^4 + 4 t^2 \sin^2(\sqrt{\pi n}a) (U^2 + \gamma_1^2)
\right]^{1/2}
\right)^{1/2}.
\end{aligned}
\end{equation} 
Both $V$ and $U$ are further used in the effective Hamiltonian onsite energy. %$\boldsymbol{\varepsilon}$. 

\section{Model functions for the gate capacitance}
As described in the main text, the ring-shaped gate is given by a model function
$C_\mathrm{ring}(x,y) =  C_\mathrm{ring_0} f(x,y)$, as shown in \autoref{fig:fig2}(b) of the main text, where
\begin{equation}
\begin{aligned}
\label{eq:modelC}
f(x,y) = &1- \left\{ 
1 - \left[1- \frac{1}{4} \left(\tanh \frac{r-R_\mathrm{out}}{d_\mathrm{smooth}} +1\right) \right.
\right. \\
&\left. \left(\tanh \frac{y-w/2}{d_\mathrm{smooth}} +\tanh \frac{-y-w/2}{d_\mathrm{smooth}} +2\right) \right] \\
&\left. 
\frac{1}{2}\left(\tanh \frac{r-R_\mathrm{in}}{d_\mathrm{smooth}} +1\right)
%1 - \left[1- \frac{1}{4} \left(\tanh\left(\frac{r-R_\mathrm{out}}{d_\mathrm{smooth}}\right)+1\right) \right.\right. \\
%&\left. \left(\tanh\left(\frac{y-w/2}{d_\mathrm{smooth}}\right)+\tanh\left(\frac{-y-w/2}{d_\mathrm{smooth}}\right)+2\right) \right] \\&\left. 
%\frac{1}{2}\left(\tanh\left(\frac{r-R_\mathrm{in}}{d_\mathrm{smooth}}\right)+1\right)
\right\}^{3.5} 
\end{aligned}
\end{equation} 
and for the top gate, $C_\mathrm{tg}(x,y) = C_\mathrm{tg_0} (1- f(x,y)) g(x)$, as presented in \autoref{fig:fig2}(c), with
\begin{equation}
g(x) = \left[1-\frac{1}{3} \left(\tanh \frac{x-L/2}{d_\mathrm{smooth}} + \tanh \frac{-x-L/2}{d_\mathrm{smooth}}  + 2 \right) \right]^{20} ,
\end{equation} 
where the multiplication by $g(x)$ is for cropping the top gate capacitance at $x=\pm L/2$ [see \autoref{fig:fig2}(c)]. In (\autoref{eq:modelC}), $r=\sqrt{x^2+y^2}$, and 
%Here, 
we assume $C_\mathrm{ring_0}=C_\mathrm{bg}/0.77$ and $C_\mathrm{tg_0}=C_\mathrm{bg}/0.99$ in accordance with \cite{Iwakiri2022}.
$R_\mathrm{in}$ and $R_\mathrm{out}$ are the inner and outer radius of the ring, respectively, %\textcolor{red}{as marked in \autoref{fig:fig2}(b)}, 
$w$ is the width of the channel, and $L$ is the length of the top gate.

%\iffalse
%\fi

\end{document}